\documentclass[10pt,journal,cspaper,compsoc]{IEEEtran}
\pdfoutput=1

\usepackage{graphicx}
\usepackage{color}

\ifCLASSOPTIONcompsoc
\else
\fi
\ifCLASSINFOpdf
\else
\fi
\begin{document}
%
\title{A Practical, Secure, and Verifiable Cloud Computing for Mobile Systems}
%
%
%
%

\author{Sriram~N.~Premnath, 
        Zygmunt~J.~Haas,~\IEEEmembership{Fellow,~IEEE}
\IEEEcompsocitemizethanks{\IEEEcompsocthanksitem S.N. Premnath is with the School
of Electrical and Computer Engineering, Cornell University, Ithaca,
NY, 14853.\protect\\
E-mail: sriram.np@cornell.edu
\IEEEcompsocthanksitem Z.J. Haas is with the School of Electrical and Computer Engineering, Cornell University, Ithaca, NY, 14853, and Department of Computer Science, University of Texas at Dallas, Richardson, TX 75080.\protect\\
E-mail: haas@ece.cornell.edu}
\thanks{}}

%
%

\markboth{IEEE TRANSACTIONS ON CLOUD COMPUTING,~Vol.~x, No.~x, January~20xx}%
{Premnath \MakeLowercase{\textit{et al.}}: A Practical, Secure, and Verifiable Cloud Computing for Mobile Systems}
%


\IEEEcompsoctitleabstractindextext{%
\begin{abstract}
Cloud computing systems, in which clients rent and share computing resources of third party platforms, have gained widespread use in recent years. Furthermore, cloud computing for mobile systems (i.e., systems in which the clients are mobile devices) have too been receiving considerable attention in technical literature. We propose a new method of delegating computations of resource-constrained mobile clients, in which multiple servers interact to construct an encrypted program known as garbled circuit. Next, using garbled inputs from a mobile client, another server executes this garbled circuit and returns the resulting garbled outputs. Our system assures privacy of the mobile client's data, even if the executing server chooses to collude with all but one of the other servers. We adapt the garbled circuit design of Beaver et al. and the secure multiparty computation protocol of Goldreich et al. for the purpose of building a secure cloud computing for mobile systems. Our method incorporates the novel use of the cryptographically secure pseudo random number generator of Blum et al. that enables the mobile client to efficiently retrieve the result of the computation, as well as to verify that the evaluator actually performed the computation. We analyze the server-side and client-side complexity of our system. Using real-world data, we evaluate our system for a privacy preserving search application that locates the nearest bank/ATM from the mobile client. We also measure the time taken to construct and evaluate the garbled circuit
for varying number of servers, demonstrating the feasibility of our secure and verifiable cloud computing for mobile systems.\vspace{-2 mm}
\end{abstract}

}

\maketitle

\IEEEdisplaynotcompsoctitleabstractindextext

%
\IEEEpeerreviewmaketitle

\section{Introduction}
\label{sec_introduction}

\IEEEPARstart{C}{loud} computing systems, in which the clients rent and share computing resources of third party platforms such as Amazon Elastic Cloud, Microsoft Azure, etc., have gained widespread use in recent years. Provisioned with a large pool of hardware and software resources, these cloud computing systems enable clients to perform computations on a vast amount of data without setting up their own infrastructure \cite{armbrust10}. However, providing the cloud service provider with the client data in {\em plaintext form} to carry out the computations will result in complete loss of data privacy.

Homomorphic encryption \cite{rivest78} is an approach to tackle the problem of preserving data privacy, which can allow the cloud service providers to perform specific computations directly on the encrypted client data, without requiring private decryption keys. Recently, fully homomorphic encryption (FHE) schemes (e.g., Gentry et al. \cite{gentry10}) have been proposed, which enable performing any arbitrary computation on encrypted data. {\em However, FHE schemes are currently impractical for mobile cloud computing applications due to \underline{extremely large cipher text size}}. For instance, to achieve $128$-bit security, the client is required to exchange a few Giga bytes of ciphertext with the cloud server, for each bit of the plain text message \cite{gentry10}. {\em Thus, there is a need for a more efficient alternative suitable for mobile systems.}

Yao's garbled circuits approach \cite{yao82,yao86}, which we consider in our work, is a potential alternative to FHE schemes that can drastically reduce the ciphertext size. Any computation can be represented using a Boolean circuit, for which, there exists a corresponding garbled circuit \cite{yao82,yao86,goldreich04,goldreich87}. Each gate in a garbled circuit can be unlocked using a pair of input {\em wire keys} that correspond to the underlying plaintext bits; and the association between the wire keys and the plaintext bits is kept secret from the cloud server that performs the computation. Unlocking a gate using a pair of input wire keys reveals an output wire key, which, in turn, serves as an input wire key for unlocking the subsequent gate in the next level of the circuit. Thus, garbled circuits can enable {\em oblivious evaluation} of any arbitrary function, expressible as a Boolean circuit, on a third-party cloud server.

While garbled circuits preserve the privacy of client data, they are, however, one time programs -- using the same version of the circuit more than once compromises the garbled circuit and reveals to an adversarial evaluator whether the semantics have changed or remained the same for a set of input and output wires between successive evaluations. Expecting the client to create a new version of the garbled circuit for each evaluation, however, is an unreasonable solution, since creating a garbled circuit is at least as expensive as evaluating the underlying Boolean circuit! {\em Thus, in contrast to FHE schemes such as that of Gentry \cite{gentry10}, that can directly delegate the desired computation to the cloud servers, a scheme using garbled circuits, presents the additional challenge of efficiently delegating to the cloud servers the creation of garbled circuit.}

We propose a new method, in which whenever the client needs to perform a computation, the client employs a number of cloud servers to create a new version of the garbled circuit in a distributed manner. Each server generates a set of private input bits using unique seed value from the client and interacts with all the other servers to create a new garbled circuit, which is a function of the private input bits of all the servers. Essentially, the servers engage in a secure multiparty computation protocol (e.g., Goldreich et al. \cite{goldreich04,goldreich87}) to construct the desired garbled circuit without revealing their private inputs to one another. Once a new version of the garbled circuit is created using multiple servers, the client delegates the evaluation to an arbitrary server in the cloud. The resulting version of the garbled circuit, the garbled inputs that can unlock the circuit, and the corresponding garbled outputs, remain unrecognizable to the evaluator, even if it chooses to collude with any strict-subset of servers that participated in the creation of the garbled circuit.

Our proposed system is designed to readily exploit the real-world asymmetry that exists between typical mobile clients and cloud servers -- while the mobile client-s are resource constrained, the cloud servers, on the other hand, are sufficiently provisioned to perform numerous intensive computation and communication tasks. To achieve secure and verifiable computing capability, our system requires very little computation and communication involvement from the mobile client beyond the generation and exchange of \underline{\em compact cipher text messages}. However, using significantly larger resources available to them, the cloud servers can efficiently generate and exchange a large volume of random bits necessary for carrying out the delegated computation. Thus, our proposed scheme is very suitable for mobile environments\footnote{While our proposed system is especially beneficial for clients in a mobile environment, due to compact cipher text messages, it is also suitable for clients in other environments that need to delegate its computations to the cloud servers in a secure manner.}.

We adapt the garbled circuit design of Beaver, Micali, Rogaway (BMR \cite{beaver90,rogaway91}), and the secure multiparty computation protocol of Goldreich et al. \cite{goldreich04,goldreich87} to suit them for the purpose of building a secure cloud computing system. To facilitate the construction of the garbled circuit, and also to enable the client to {\em efficiently retrieve and verify the result of the computation}, our method incorporates the novel use of the cryptographically secure pseudo random number generator of Blum, Blum, Shub \cite{blum86,schneier95}, whose strength relies on the computational difficulty of factorizing large numbers into primes. {\em Our proposed system enables the client to efficiently verify that the evaluator actually and fully performed the requested computation.}

Our major contributions in this work include the following: (i) we design a secure mobile cloud computing system using multiple servers that enables the client to delegate any arbitrary computation, (ii) our system assures the privacy of the client input and the result of the computation, even if the evaluating server colludes with all but one of the servers that created the garbled circuit, (iii) our system enables the client to efficiently retrieve/recover the result of the computation and to verify whether the evaluator actually performed the computation, (iv) we present an analysis of the server-side and client-side complexity of our proposed scheme. Our findings show that in comparison to Gentry's FHE scheme, our scheme uses very small cipher text messages suitable for mobile clients, (v) using real-world data, we evaluate our system for a privacy preserving search application that locates the nearest bank/ATM from the mobile client, and (vi) we measure the time taken to construct and evaluate the garbled circuit for varying number of servers, demonstrating the feasibility of our system.\vspace{-2 mm}

\section{A High-level Overview of our\\ System}

In the proposed system, the client employs a set of $(n+2)$ servers, $\{p_1, p_2, \ldots, p_n, p_c, p_e\}$. Initially, the client sends a description of the desired computation (such as addition of two numbers, computation of hamming distance between two bit sequences, etc.), and a unique seed value $s_i$ to each server $p_i, (1\le i\le n)$. Each of these $n$ servers first creates (or retrieves from its repository, if available already) a Boolean circuit ($B$) that corresponds to the requested computation. Using the unique seed value $s_i$, each server $p_i, (1\le i\le n)$ generates a private pseudorandom bit sequence whose length is proportional to the total number of wires in the Boolean circuit ($B$). Then, using the private pseudorandom bit sequences and the Boolean circuit ($B$) as inputs, these $n$ servers interact with one another, while performing some local computations, according to a secure multiparty computation protocol, to create their shares ($GC_i, (1\le i\le n)$) for an one-time program called garbled circuit.

Once the shares for the garbled circuit are created, the client requests each server, $p_i, (1\le i\le n)$, to send its share, $GC_i$, to the server $p_c$. Performing an XOR operation on these shares, the server $p_c$ creates the desired circuit, $GC = GC_1\oplus GC_2\oplus \ldots \oplus GC_n$. Subsequently, the client instructs the server $p_c$ to send the garbled circuit $GC$ to another server $p_e$ for evaluation.

Now, using the unique seed values $s_i, (1\le i\le n)$, the client generates on its own garbled input values for each input wire in the circuit and sends them to the server $p_e$ for evaluation. Using these garbled inputs, the server $p_e$ unlocks the gates in the first level of the circuit to obtain the corresponding garbled outputs, which, in turn, unlocks the gates in the second level of the circuit, and so on. In this manner, the server $p_e$ unlocks all the gates in the circuit, obtains the garbled outputs of the circuit, and sends them to the client. The client now converts these garbled output values into plaintext bits to recover the result of the desired computation.

\begin{figure}[t]
\centering
\includegraphics[width=3.25in]{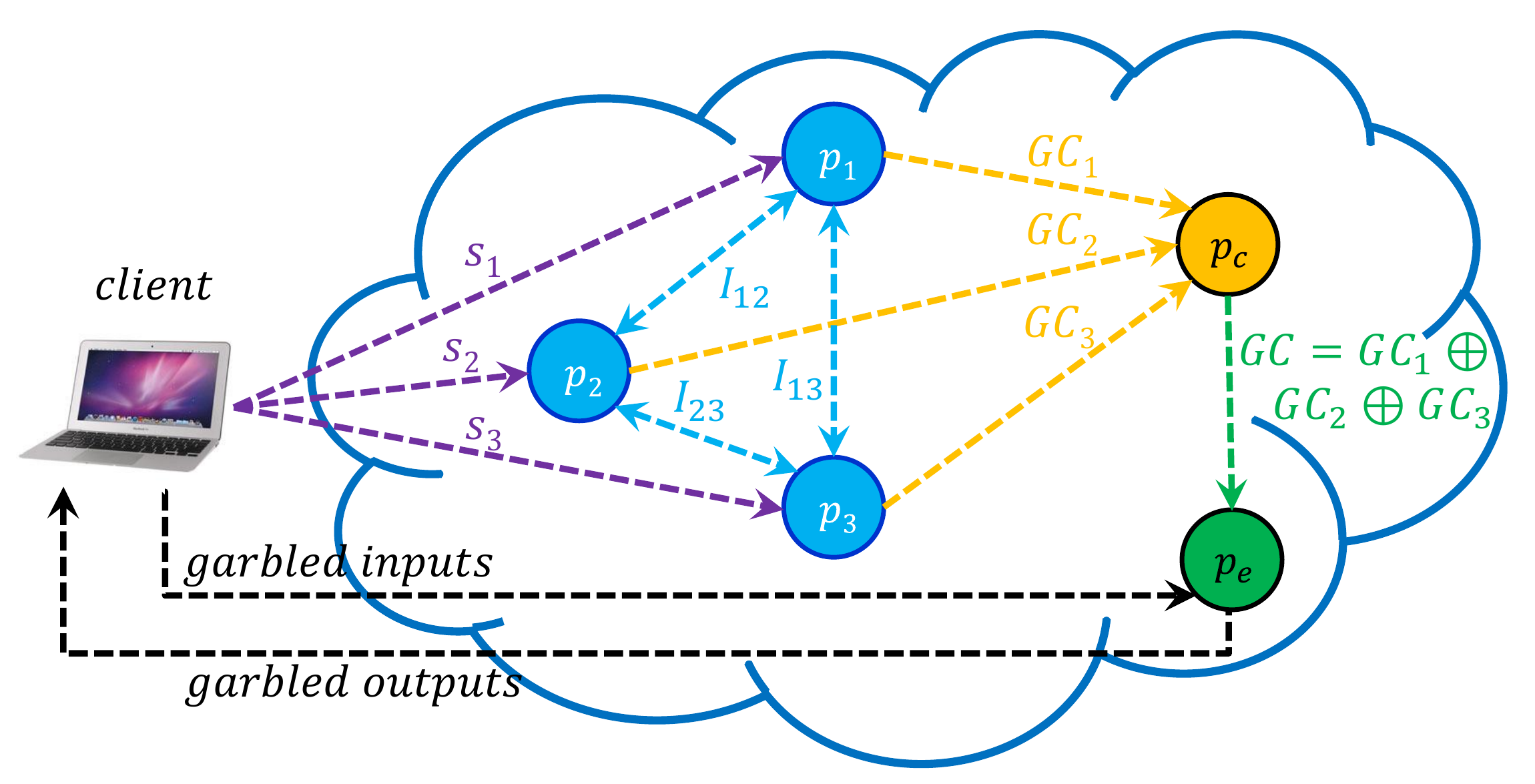}
\vspace{-3 mm}
\caption{Our secure cloud computing model with $(n+2)=5$ servers. 1.) Client sends unique seed value, $s_i$, to each $p_i, (1\le i\le 3)$; 2.) $p_1,p_2,p_3$ interact ($I_{ij}, 1\le i<j\le 3$) to construct shares of the garbled circuit $GC$; 3.) Each $p_i$ sends its share ($GC_i$) to $p_c$; 4.) $p_c$ computes $GC=GC_1\oplus GC_2\oplus GC_3$, and sends it to $p_e$; 5.) Client generates garbled inputs, and sends them to $p_e$; 6.) $p_e$ evaluates $GC$, and sends the garbled outputs to the client.\vspace{-4 mm}}
\label{fig_our_secure_cloud_computing_Model}
\end{figure}

Fig.~\ref{fig_our_secure_cloud_computing_Model} depicts an overview of our secure cloud computing system model with $(n+2)=5$ servers.

\subsection{Our Adversary Model}
\label{sec_adversary_model}

We assume the existence of a secure communication channel between the client and each of the $(n+2)$ servers, $\{p_1, p_2, \ldots, p_n, p_c, p_e\}$, for sending unique seed values for pseudorandom bit generation, identity of the other servers, etc. We assume that all pairs of communicating servers authenticate one another. We assume a very capable adversary, where the evaluator $p_e$ may individually collude with any proper subset of the $n$ servers, $\{p_1, p_2, \ldots, p_n\}$, and still remain unable to determine the semantics of any garbled value that the evaluator observes during evaluation. Thus, our adversary model depicts a very realistic scenario -- where the client may be certain that some (however, not all) of the parties are corrupt, however, it is uncertain which of the parties are corrupt. If any adversarial party attempts to eavesdrop and analyze the set of all message exchanges between different parties, and also analyze the set of all the messages that it has legitimately received from the other parties, it still cannot determine the shares of the other parties, or the semantics of the garbled value pairs that are assigned to each wire in the circuit. Further, if the evaluator, $p_e$, returns arbitrary numbers as outputs to the client, the client can detect this efficiently. In our model, a new garbled circuit is created for every evaluation. This prevents an adversarial evaluator from determining the set of inputs and outputs that have changed or remained the same between different evaluations.

\subsection{Main Characteristics of our system}
\label{sec_features_of_our_secure_cloud_computing_model}
We highlight some of the main features of our secure cloud computing system in this subsection.
\begin{enumerate}
    \item{{\em Offloaded Computation}: The client delegates the intensive computational tasks to the cloud servers of creating and evaluating the garbled circuit. The client only chooses the cloud servers, provides them with unique seed values, generates garbled inputs during evaluation, and interprets garbled outputs returned by the evaluator.}

    \item{{\em Compact cipher text}: While Gentry's scheme has an extremely large cipher text size (on the order of several Gigabits), the cipher text size can be as small as a few hundred bits with our scheme, as we have shown in Section~\ref{}. Thus, our proposed method is far more practical for cloud computing in mobile systems in comparison to FHE schemes.}

    \item{{\em Decoupling}: The process of creating the garbled circuit is decoupled from the process of evaluating the garbled circuit. While the servers, $p_i, (1\le i\le n)$, interact with one another for creating the garbled circuit, the server $p_e$ evaluates the garbled circuit, {\em independently}.}

    \item{{\em Precomputation of garbled circuits}: Since evaluation of the garbled circuit requires no interaction among the servers, if several versions of the garbled circuit for a given computation are precomputed and stored at the evaluator, {\em in advance}, then it can readily carry out the requested computation. {\em Thus, the client will only incur the relatively short time taken to evaluate the garbled circuit. In other words, precomputation will drastically improve the response time for the client.}}

    \item{{\em Collusion Resistance}: To interpret any garbled value, the evaluator, $p_e$, would need to collude with all the $n$ servers, $p_i, (1\le i\le n)$. Thus, even if $(n-1)$ out of the $n$ servers are corrupt and collude with the evaluator, the privacy of the client's inputs and the result of the computation are still preserved.}

    \item{{\em Verification of outputs}: The client has the ability to verify that the evaluator actually carried out the requested computation. 
}
\end{enumerate}

\section{Background}
\label{sec_background}
We briefly describe the construction and evaluation of Yao's garbled circuits \cite{yao82,yao86}, as well as the oblivious transfer protocols of Naor and Pinkas \cite{naor01,naor05}.

\subsection{Yao's Garbled Circuit}

\begin{figure}[ht]
\centering
\includegraphics[width=2.4in]{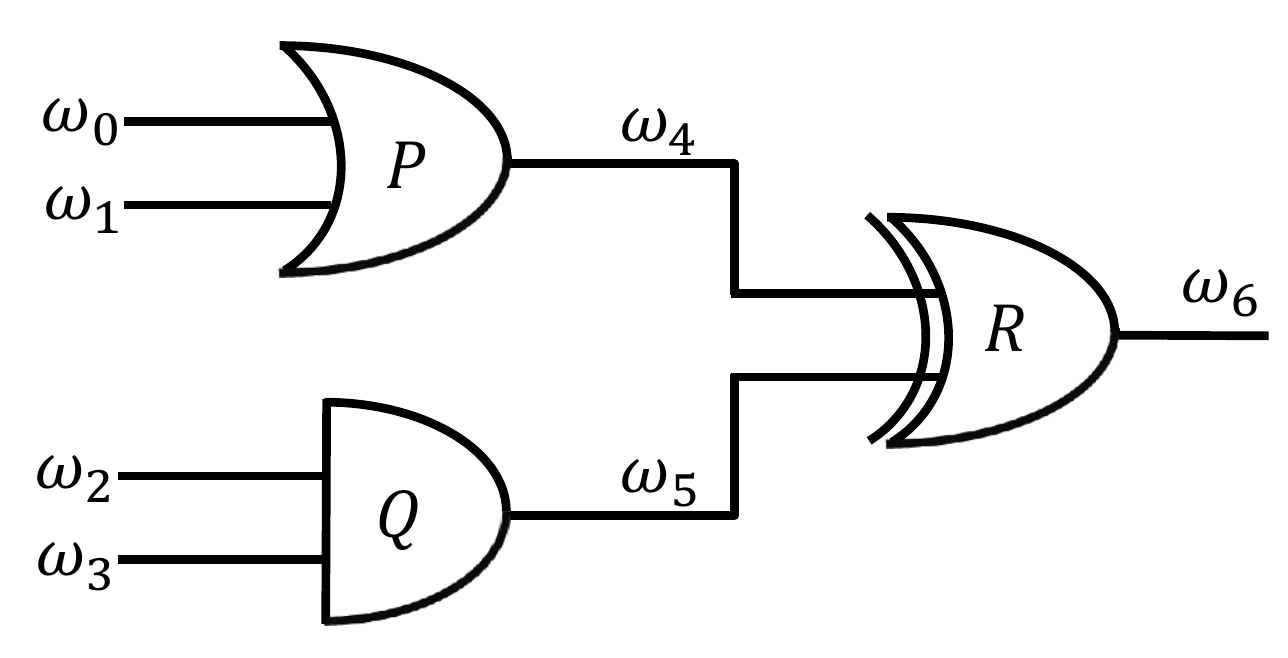}
\vspace{-3 mm}
\caption{A circuit with three gates, $P,Q,R$, and seven wires, $\omega_i, (0\le i\le 6)$.\vspace{-3 mm}}
\label{fig_sample_circuit}
\end{figure}

Each wire in the circuit is associated with a pair of keys known as {\em garbled values} that correspond to the underlying binary values. For example, the circuit in Figure~\ref{fig_sample_circuit} has seven wires, $\omega_i, (0\le i\le 6)$, and three gates, $P,Q,R$, denoting OR, AND, XOR gates respectively. Keys $\omega_i^0$, $\omega_i^1$ represent the garbled values corresponding to binary values $0$, $1$ respectively on the wire $\omega_i$.

\begin{table}[!t]
\renewcommand{\arraystretch}{1.3}
\caption{Garbled Tables for Gates $P,Q,R$.}
\label{table_garbled_tables}
\centering
\begin{tabular}{||c||c||c||}
\hline
\bfseries P & \bfseries Q & \bfseries R\\
\hline
$E_{\omega_0^0}[E_{\omega_1^1}[\omega_4^1]$ & $E_{\omega_2^0}[E_{\omega_3^1}[\omega_5^0]$ & $E_{\omega_4^1}[E_{\omega_5^1}[\omega_6^0]$\\
$E_{\omega_0^0}[E_{\omega_1^0}[\omega_4^0]$ & $E_{\omega_2^1}[E_{\omega_3^0}[\omega_5^0]$ & $E_{\omega_4^0}[E_{\omega_5^0}[\omega_6^0]$\\
$E_{\omega_0^1}[E_{\omega_1^1}[\omega_4^1]$ & $E_{\omega_2^0}[E_{\omega_3^0}[\omega_5^0]$ & $E_{\omega_4^1}[E_{\omega_5^0}[\omega_6^1]$\\
$E_{\omega_0^1}[E_{\omega_1^0}[\omega_4^1]$ & $E_{\omega_2^1}[E_{\omega_3^1}[\omega_5^1]$ & $E_{\omega_4^0}[E_{\omega_5^1}[\omega_6^1]$\\
\hline
\end{tabular}
\vspace{-3 mm}
\end{table}

Each gate in the circuit, is associated with a list of four values, in a random order, known as {\em garbled table}. Table~\ref{table_garbled_tables} shows the garbled tables for the gates $P,Q,R$ of Fig.~\ref{fig_sample_circuit}. Let $x,y\in\{0,1\}$. Let $E_{k}[v]$ denote the encryption of $v$ using $k$ as the key. Then, each entry in the garble table for $P$ is of the form, $E_{\omega_0^x}[E_{\omega_1^y}[\omega_4^{x|y}]$. Similarly, each entry in the garble table for $Q$ and $R$ are of the forms, $E_{\omega_2^x}[E_{\omega_3^y}[\omega_5^{x.y}]$ and $E_{\omega_4^x}[E_{\omega_5^y}[\omega_6^{x\oplus y}]$, respectively.

Suppose that the client wishes to delegate the computation of Fig.~\ref{fig_sample_circuit}, i.e., $((a|b)\oplus (c.d))$, to a server in the cloud. The server is provided with a description of the circuit (Fig.~\ref{fig_sample_circuit}) along with the set of the garbled tables (Table~\ref{table_garbled_tables}), which together represents a {\em garbled circuit}. However, the client keeps the mapping between the garbled values and the underlying binary values as secret. For example, to evaluate the circuit with inputs $a=1,b=0,c=0,d=1$, the client provides the set of garbled inputs, $\omega_0^1,\omega_1^0,\omega_2^0,\omega_3^1$, to the cloud server.

Now, assume that there exists a mechanism to determine whether a value is decrypted correctly; for example, through zero-padding. Using $\omega_0^1,\omega_1^0$ as keys, the server attempts to decrypt all the four entries in the garbled table for gate $P$; however, only the fourth entry will decrypt correctly to reveal the garbled output $\omega_4^1$. Similarly, on using $\omega_2^0,\omega_3^1$ as keys, the first entry in the garbled table for gate $Q$ reveals the garbled output $\omega_5^0$. Finally, on using $\omega_4^1,\omega_5^0$ as keys, the third entry in the garbled table for gate $R$ reveals the garbled output $\omega_6^1$. Thus, the server can perform an {\em oblivious evaluation} of the garbled circuit and return the result of the computation $\omega_6^1$ to the client. Using the secret mapping, the client can determine that the garbled value $\omega_6^1$ corresponds to the binary value $1$.

In our work, we use an alternative garbled circuit design from Beaver, Micali, Rogaway (BMR \cite{beaver90,rogaway91}), and adapt it, as we describe in Section~\ref{sec_our_cloud_computing_model}, for the purpose of building a secure cloud computing system.

\subsection{1-out-of-2 Oblivious Transfer}
\label{sec_1_out_of_2_OT}
There are two parties, a sender and a chooser. The sender holds two messages, $M_0,M_1$, and the chooser holds a choice bit, $\sigma \in \{0,1\}$. At the end of the 1-out-of-2 oblivious transfer (OT) protocol, the chooser learns $M_\sigma$ only, while the sender learns nothing.

Let $p=2q+1$ denote a safe prime number; i.e., $q$ is also a prime number. Let $Z_p^*=\{1,2,3,4,\ldots ,(p-1)\}$, which denotes the set of integers that are relatively prime to $p$. Let $G$ denote a subgroup of $Z_p^*$, where $|G|=q$. Let $g$ denote the generator for $G$.

The sender randomly chooses an element, $C\in G$, and sends it to the chooser. Note that the discrete logarithm of $C$ is unknown to the chooser. The chooser randomly selects an integer, $k, (1\le k\le q)$, and sets $PK_\sigma=g^k\; mod\; p$, and $PK_{1-\sigma}=C\times(PK_\sigma)^{-1}\; mod\; p$. The chooser sends $PK_0$ to the sender. Note that $PK_0$ does not reveal the choice bit $\sigma$ to the sender.

The sender calculates $PK_1=C\times (PK_0)^{-1}\; mod\; p$ on its own, and randomly chooses two elements, $r_0,r_1\in G$. Let $h(x)$ denote the output of the hash function (e.g., SHA) on input $x$. Let $E_i$ denote the encryption of $M_i$, $\forall i\in\{0,1\}$. Then, the sender calculates $E_i=[(g^{r_i}\; mod\; p),(h(PK_i^{r_i}\; mod\; p)\oplus M_i)]$, and sends both $E_0$, $E_1$ to the chooser.

The chooser decrypts $E_\sigma$ to obtain $M_\sigma$ as follows. Let $l_1=g^{r_\sigma}\; mod\; p$ and $l_2=h(PK_\sigma^{r_\sigma}\; mod\; p)\oplus M_\sigma$ denote the first and second numbers respectively in $E_\sigma$. The chooser calculates $M_\sigma$ using the relation, $M_\sigma=h(l_1^k\; mod\; p)\oplus l_2$. Note that since the discrete logarithm of $C$, and hence $PK_{1-\sigma}$, is unknown to the chooser, it cannot retrieve $M_{1-\sigma}$ from $E_{1-\sigma}$.

\subsection{1-out-of-4 Oblivious Transfer}
\label{sec_1_out_of_4_OT}

There are two parties, a sender and a chooser. The sender holds four messages, $M_{00}$, $M_{01}$, $M_{10}$, $M_{11}$, and the chooser holds two choice bits, $\sigma_1,\sigma_2$. At the end of the 1-out-of-4 oblivious transfer (OT) protocol, the chooser learns $M_{\sigma_1 \sigma_2}$ only, while the sender learns nothing.

The sender randomly generates two pairs of keys, $(L_0,L_1),(R_0,R_1)$, and computes the encryptions of $M_{00}$, $M_{01}$, $M_{10}$, $M_{11}$ as follows. Let $F_k (x)$ denote the output of a pseudorandom function such as AES-128, that is keyed using $k$ on the input $x$. Let $E_{ij}$ denote the encryption of $M_{ij}$, $\forall i,j\in\{0,1\}$. Then, $E_{ij}=M_{ij}\oplus F_{L_i}(2i+j+1)\oplus F_{R_j}(2i+j+1)$.

The sender and the chooser engage in 1-out-of-2 OT twice. In the first 1-out-of-2 OT, the sender holds two messages, $L_0,L_1$, and the chooser holds the choice bit, $\sigma_1$; at the end of this OT, the chooser obtains $L_{\sigma_1}$. In the second 1-out-of-2 OT, the sender holds two messages, $R_0,R_1$, and the chooser holds the choice bit, $\sigma_2$; at the end of this OT, the chooser obtains $R_{\sigma_2}$.

Now, the sender sends all the four encryptions, $E_{00}$, $E_{01}$, $E_{10}$, $E_{11}$, to the chooser. Using $L_{\sigma_1}$, $R_{\sigma_2}$, the chooser decrypts $E_{\sigma_1 \sigma_2}$ to obtain $M_{\sigma_1 \sigma_2}$, as $M_{\sigma_1 \sigma_2}=E_{\sigma_1 \sigma_2}\oplus F_{L_{\sigma_1}}(2\sigma_1+\sigma_2+1)\oplus F_{R_{\sigma_2}}(2\sigma_1+\sigma_2+1)$.

\section{Secure and Verifiable Cloud Computing for Mobile Systems}
\label{sec_our_cloud_computing_model}

In Section~\ref{sec_garbled_circuit_BMR}, we present the construction of BMR garbled circuit \cite{beaver90,rogaway91} using $n$ servers through the secure multiparty computation protocol of Goldreich et al. \cite{goldreich04,goldreich87}. In Section~\ref{sec_key_changes_to_BMR_Goldreich}, we highlight how we adapt the protocol of Goldreich and the garbled circuit design of BMR, in order to suit them for our secure cloud computing model. In our model, each server $p_i, (1\le i\le n)$, generates shares of garbled values using cryptographically secure pseudorandom number generation method of Blum, Blum, Shub \cite{blum86,schneier95}. In Section~\ref{sec_output_verification}, we present our method of how the client efficiently recovers the result of the delegated computation, as well as how the client verifies that the evaluator in fact carried out the computation. We summarize our secure cloud computing model in Section~\ref{sec_our_secure_cloud_computing_model_summary}.

\subsection{Construction and Evaluation of Garbled Circuits}
\label{sec_garbled_circuit_BMR}

\subsubsection{Construction of the garbled circuit, $GC$}
\label{sec_garbled_circuit_BMR_construction}

\underline{\em Garbled Value Pairs:} Each wire in the circuit is associated with a pair of garbled values representing the underlying plaintext bits $0$ and $1$. Let $A$ denote a specific gate in the circuit, whose two input wires are $x,y$; and whose output wire is $z$. Let $(\alpha_0, \alpha_1)$, $(\beta_0, \beta_1)$ and $(\gamma_0, \gamma_1)$ denote the pair of garbled values associated with the wires $x$, $y$ and $z$, respectively. Note that $LSB(\alpha_0) = LSB(\beta_0) = LSB(\gamma_0) = 0$ and $LSB(\alpha_1) = LSB(\beta_1) = LSB(\gamma_1) = 1$.

Each garbled value is $(nk+1)$ bits long, where $n$ denotes the number of servers and $k$ denotes the security parameter. Essentially, each garbled value is a concatenation of shares from the $n$ servers. Let $a, b ,c \in\{0, 1\}$. Then the garbled values are expressed as follows: $\alpha_{a} = \alpha_{a1} || \alpha_{a2} || \alpha_{a3} || \ldots || \alpha_{an} ||a$; $\beta_{b} = \beta_{b1} || \beta_{b2} || \beta_{b3} || \ldots || \beta_{bn} || b$; $\gamma_{c} = \gamma_{c1} || \gamma_{c2} || \gamma_{c3} || \ldots ||\gamma_{cn} || c$; where $\alpha_{ai}$, $\beta_{bi}$, $\gamma_{ci}$ are shares of server $p_i, (1\le i\le n)$.

\underline{\em $\lambda$ Value:} Each wire in the circuit is also associated with a $1$-bit $\lambda$ value that determines the semantics for the pair of garbled values. Specifically, the garbled value whose $LSB = b$ represents the underlying plaintext bit $(b\oplus \lambda)$.

\underline{\em Collusion Resistance:} Let $\lambda_x, \lambda_y, \lambda_z$ denote the $\lambda$ values for the wires $x,y,z$ respectively. Then, $\lambda_x = \bigoplus_{i = 1}^{n}\lambda_{xi}$, $\lambda_y = \bigoplus_{i = 1}^{n}\lambda_{yi}$, $\lambda_z = \bigoplus_{i = 1}^{n}\lambda_{zi}$, where $\lambda_{xi}, \lambda_{yi}, \lambda_{zi}$ are shares of server $p_i, (1\le i\le n)$. Note that the $\lambda$ value of each wire is unknown to any individual server. Consequently, the evaluator of the garbled circuit must collude with all the $n$ servers to interpret the garbled values.

\underline{\em Garbled Table:} Each gate, $A$, in the circuit, is associated with an ordered list of four values, $[A_{00}, A_{01}, A_{10}, A_{11}]$, which represents the garbled table for gate $A$. Let $\otimes \in \{XOR,AND\}$ denote the binary operation of gate $A$. Then, the value of one specific entry, $A_{ab} = \gamma_{[((\lambda_x \oplus a) \otimes (\lambda_y \oplus b)) \oplus \lambda_z]} \oplus [G_b(\alpha_{a1})\oplus G_b(\alpha_{a2})\oplus \ldots \oplus G_b(\alpha_{an})]\oplus [G_a(\beta_{b1})\oplus G_a(\beta_{b2})\oplus \ldots \oplus G_a(\beta_{bn})]$, where $G_a$ and $G_b$ are pseudorandom functions that expand $k$ bits into $(nk+1)$ bits. Specifically, let $G$ denote a pseudorandom generator, which on providing a $k$-bit input seed, outputs a sequence of $(2nk+2)$ bits, i.e., if $|s|=k$, then $|G(s)|=(2nk+2)$. $G$ may represent the output of AES block cipher in output feedback mode, for example. Then, $G_0(s)$ and $G_1(s)$ denote the first and last $(nk+1)$ bits of $G(s)$ respectively.

To compute a garbled table entry $A_{ab}$, such as the one shown above, the $n$ servers use the secure multiparty computation protocol of Goldreich \cite{goldreich04,goldreich87} (Section~\ref{sec_secure_multiparty_construction_Goldreich}), where $f(x_1, x_2, \ldots, x_n) = A_{ab}$, and for each server, $p_i, (1\le i \le n)$, its private input, $x_i = [\lambda_{xi}, \lambda_{yi}, \lambda_{zi}, G_b(\alpha_{ai}), G_a(\beta_{bi}), \gamma_{0i}, \gamma_{1i}]$ is a vector of length $m=(3+2(nk+1)+2k)$ bits. In this manner, the $n$ servers jointly compute each entry in the garbled table for each gate in the circuit.

\subsubsection{Secure multiparty computation of an entry, $A_{ab}$}
\label{sec_secure_multiparty_construction_Goldreich}

Assume that $n$ parties need to compute the value of an arbitrary function of their private inputs, namely $f(x_1, x_2, \ldots, x_n)$ without revealing their private inputs to one another. Assume that the function $f(x_1, x_2, \ldots, x_n)$ is expressed as a Boolean circuit ($B'$)\footnote{Boolean circuit $B'$ is different from Boolean circuit $B$. While $B$ is a circuit that corresponds to the computation requested by the client (e.g., addition of two numbers), $B'$ is a circuit that creates the entries such as $A_{ab}$ in the garbled tables of the garbled circuit $GC$.} using a set of XOR and AND gates.

We briefly describe the secure multipary computation protocol of Goldreich \cite{goldreich04,goldreich87} as follows. For each wire in the Boolean circuit, the actual binary value corresponds to the XOR-sum of shares of all the $n$ parties.

Evaluation of each XOR gate in the circuit is carried out locally. Specifically, each party merely performs an XOR operation over its shares for the two input wires to obtain its share for the output wire.

Evaluation of each AND gate in the circuit, on the other hand, requires communication between all pairs of parties. For the two inputs wires to the AND gate, let $a_i, b_i$ denote the shares of party $p_i$; and let $a_j, b_j$ denote the shares of party $p_j$. Then, the XOR-sum of the shares for the output wire of the AND gate is expressed as follows:

$(\bigoplus_{i=1}^{n} a_i).(\bigoplus_{i=1}^{n} b_i)=[\bigoplus_{1\le i<j\le n} ((a_i\oplus a_j).(b_i\oplus b_j))]\\\bigoplus_{i=1}^{n} ((a_i.b_i).I)$, where $I=n$ $mod$ $2$.

Each party $p_i$ locally computes $((a_i.b_i).I)$; and the computation of each {\em partial-product}, $((a_i\oplus a_j).(b_i\oplus b_j))$, is accomplished using 1-out-of-4 oblivious transfer (OT) between $p_i$ and $p_j$, such that no party reveals its shares to the other party \cite{goldreich04,goldreich87}.

Following the above procedure, the $n$ parties evaluate every gate in the circuit. Thus, in the end, for the BMR protocol, as we have described above,  each server $p_i, (1\le i\le n)$, obtains the share, $(f(x_1, x_2, \ldots, x_n))_i = (A_{ab})_i$, such that $f(x_1, x_2, \ldots, x_n)=A_{ab}=\bigoplus_{i=1}^{n}(A_{ab})_i$.

\subsubsection{Evaluation of the garbled circuit, $GC$}
\label{sec_garbled_circuit_BMR_evaluation}

The garbled table for each gate, $A$, in the circuit is an ordered list of four values, $[A_{00}, A_{01}, A_{10}, A_{11}]$.

Let $\alpha, \beta$ denote the garbled values for the two input wires of a gate during evaluation. Let $a, b$ denote the LSB values of $\alpha, \beta$ respectively. Then, the garbled value for the output wire, $\gamma$, is recovered using $\alpha, \beta, A_{ab}$, as shown in the two-step process below:

\begin{enumerate}
    \item{split the most significant $nk$ bits of $\alpha$ into $n$ parts, $\alpha_1, \alpha_2, \alpha_3, \ldots, \alpha_n$, each with $k$ bits; similarly, split the most significant $nk$ bits of $\beta$ into $n$ parts, $\beta_1, \beta_2, \beta_3, \ldots, \beta_n$, each with $k$ bits; i.e., $|\alpha_i| = |\beta_i| = k$, where $1\le i\le n$.}

    \item{compute $\gamma = [G_b(\alpha_1)\oplus G_b(\alpha_2)\oplus \ldots \oplus G_b(\alpha_n)]\\\oplus [G_a(\beta_1)\oplus G_a(\beta_2)\oplus \ldots \oplus G_a(\beta_n)]\oplus A_{ab}$.}
\end{enumerate}

Thus, the garbled output for any gate in the circuit can be computed using the garbled table and the two garbled inputs to the gate. Note that while the construction of the garbled circuit requires interaction among all the $n$ parties, $p_i,(1\le i\le n)$, the server $p_e$ can perform the evaluation {\em independently}.

\subsection{Secure and Verifiable Cloud Computing through Secure Multiparty Computation}
\label{sec_key_changes_to_BMR_Goldreich}

In a secure multiparty computation protocol, multiple parties hold private inputs, and receive the result of the computation. However, in our proposed secure cloud computing system, while multiple parties participate in the creation of garbled circuits, only the client holds private inputs and obtains the result of the computation in garbled form. {\em Therefore, we adapt the protocols of Goldreich and BMR in a number of ways, as we discuss in this section, to build an efficient, secure cloud computing system, that also enables the client to easily verify the outputs of the computation.}

First, note that in the protocol of Goldreich \cite{goldreich04,goldreich87}, each party $p_i$ sends its share $(f(x_1, x_2, \ldots, x_n))_i$ to {\em all the other parties}. Using these shares, each party computes $f(x_1, x_2, \ldots, x_n)$ as $\bigoplus_{i = 1}^n (f(x_1, x_2, \ldots, x_n))_i$. In our secure cloud computing system, however, we require each server $p_i, (1\le i\le n)$, to send its share to {\em only one server}, $p_c$, which combines them using the XOR operation to produce entries such as $A_{ab}$ for each garbled table in the garbled circuit, $GC$.

Second, in the BMR protocol \cite{beaver90,rogaway91}, which is also a secure multiparty computation protocol, in addition to creating the garbled circuit, for evaluation, the $n$ parties also create garbled inputs using secure multiparty computation. Then, {\em each of these $n$ parties evaluates the garbled circuit and obtains the result of the computation}. In our system model, since only the client holds the inputs for the computation, it generates the corresponding garbled input for each input wire on its own using the seed values it sends to each server, $p_i, (1\le i\le n)$. Then, it sends these garbled values to server $p_e$ for evaluating the garbled circuit and obtains the result in garbled form. Note that in our model, {\em only the server $p_e$ evaluates the garbled circuit, and
that $p_e$ cannot interpret any garbled value, unless it colludes with all the $n$ servers, $p_i, (1\le i\le n)$}.

Third, in the BMR protocol \cite{beaver90,rogaway91}, the $\lambda$ value is set to zero for each output wire in the Boolean circuit. Therefore, each party evaluating the garbled circuit obtains the result of the computation in plaintext form from the LSB of the garbled output for each output wire in the circuit. In our system model, however, the $\lambda$ value for each output wire is also determined using the XOR-sum of the shares from all the $n$ servers, $p_i, (1\le i\le n)$. {\em As a consequence, result of the computation in plaintext form remains as secret for the evaluator $p_e$}.

Fourth, in the protocol of Goldreich \cite{goldreich04,goldreich87}, each party splits and shares each of its private input bits with all the other parties over pairwise secure communication channels. In our approach, we eliminate this communication using a unique seed value $s_{ik}$ that the client shares with all pairs of parties, $(p_i,p_k), (1\le i,k\le n)$. To split and share each of its $m$ private input bits $x_{ij}, (1\le j\le m)$, party $p_i$ generates $r_{kj}, (\forall k\neq i)$, using the seed value $s_{ik}$. More specifically, party $p_i$ sets its own share as $x_{ij}\bigoplus_{k=1,k\neq i}^{n}r_{kj}$, where $r_{kj}=R(s_{ik},j,gate_{id},entry_{id})$ corresponds to the output of the pseudorandom bit generator using the seed value $s_{ik}$ for the $j^{th}$ private input bit of party $p_i$, for a specific garbled table entry ($entry_{id}$) of one of the gates ($gate_{id}$) in the circuit. Likewise, party $p_k$ sets its own share as $r_{kj}=R(s_{ik},j,gate_{id},entry_{id})$. The total number of pseudorandom bits generated by each party for the protocol of Goldreich equals $2(n-1)m\times 4N_g=8(n-1)m\times N_g$, where $m=(3+2(nk+1)+2k)$, and $N_g$ denotes the total number of gates in the circuit. {\em In other words, our approach eliminates the exchange of a very large number of bits ($O(n^3 k N_g)$ bits) between the $n$ parties.}

Fifth, our novel use of the Blum, Blum, Shub pseudorandom number generator for generating garbled value shares {\em enables the client to efficiently recover and verify the outputs of the computation}. The client can detect a {\em cheating evaluator}, if it returns arbitrary values as output. We present this in a greater detail in Section~\ref{sec_output_verification}.\vspace{-1 mm}

\subsection{Recovery and Verification of Outputs}
\label{sec_output_verification}

We address the following questions in this subsection. {\em First, how does the client efficiently retrieve/recover the result of the computation without itself having to repeat the delegated computations? Second, how does the client verify that the evaluator, in fact, evaluated the garbled circuit? In other words, is it possible for the client to determine whether the evaluator returned arbitrary numbers without carrying out any computation at all, instead of the actual garbled output for each output wire?}

We can enable the client to efficiently retrieve and verify the outputs returned by the evaluator, $p_e$. To achieve this, each of the $n$ parties that participates in the creation of the garbled circuit uses the cryptographically secure Blum, Blum, Shub pseudorandom number generator \cite{blum86,schneier95}, as we have outlined below.

Let $N$ denote the product of two large prime numbers, $p,q$, which are congruent to $3$ $mod$ $4$. The client chooses a set of $n$ seed values, $\{s_1, s_2, \ldots, s_n\}$, where each seed value $s_i$ belongs to $Z_N^*$, the set of integers relatively prime to $N$. The client sends the modulus value $N$ and a unique seed value $s_i$ to each party $p_i, (1\le i\le n)$ over a secure communication channel. However, the client keeps the prime factors, $p,q$, of $N$ as a secret.

Let $b_{i,j}$ denote the $j^{th}$ bit generated by the party $p_i$. Then, $b_{i,j}=LSB(x_{i,j})$, where $x_{i,j}=x_{i,(j-1)}^2$ $mod$ $N$, and $x_{i,0}=s_i$.

Each wire $\omega$ in the circuit is associated with a pair of garbled values, $(\omega_0,\omega_1)$, and a $1$-bit $\lambda_\omega$ value. Then, $\omega_0=\omega_{01}  || \omega_{02}  || \omega_{03}  || \ldots || \omega_{0n}  || 0$, $\omega_1=\omega_{11}  || \omega_{12}  || \omega_{13}  || \ldots ||\\ \omega_{1n}  || 1$; and $\lambda_\omega=\lambda_{\omega 1}\oplus \lambda_{\omega 2}\oplus \lambda_{\omega 3} \oplus \ldots \oplus \lambda_{\omega n}$. In these three expressions, $\omega_{0i},\omega_{1i}$ and $\lambda_{\omega i}$ are shares of the party $p_i,(1\le i\le n)$. Note that $|\omega_{0i}|=|\omega_{1i} |=k$, and $|\lambda_{\omega i}|=1$.

For each wire $\omega, (0\le \omega \le W-1)$, in the circuit, each party, $p_i, (1\le i\le n)$, needs to generate $(2k+1)$ pseudo random bits, where $W$ denotes the total number of wires in the circuit. Thus, each party, $p_i$, generates a total of $(W(2k+1))$ pseudorandom bits.

Party $p_i$ generates its shares $\omega_{0i},\omega_{1i},$ and $\lambda_{\omega i}$ for wire $\omega$ as a concatenation of the $b_{i,j}$ values, where the indices $j$ belong to the range: $[(\omega(2k+1)+1),(\omega+1)(2k+1)]$. For concise notation, let $\Omega_{\omega k}=\omega(2k+1)$. Then,

$\omega_{0i}=b_{i,(\Omega_{\omega k}+1)} || b_{i,(\Omega_{\omega k}+2)} || \ldots || b_{i,(\Omega_{\omega k}+k)}$,

$\omega_{1i}=b_{i,(\Omega_{\omega k}+k+1)} || b_{i,(\Omega_{\omega k}+k+2)} || \ldots || b_{i,(\Omega_{\omega k}+2k)}$,

$\lambda_{\omega i}=b_{i,(\Omega_{\omega k}+2k+1)}$.

\underline{\em Short-cut:} Notice that each party $p_i$ is required to compute all the previous $(j-1)$ bits before it can compute the $j^{th}$ bit. However, using its knowledge of the prime factors of $N$, i.e., $p,q$, the client can directly calculate any $x_{i,j}$ (hence, the bit $b_{i,j}$) using the relation: $x_{i,j}=x_{i,0}^{2^{j}\; mod\; C(N)}\; mod\; N$, where $C(N)$ denotes the {\em Carmichael function}, which equals the least common multiple of $(p-1)$ and $(q-1)$.

{\em Therefore, using the secret values $p,q$, the client can readily compute $\omega_0,\omega_1,$ and $\lambda_\omega$ for any output wire $\omega$ in the circuit; i.e., without having to compute $\omega_0,\omega_1,$ and $\lambda_\omega$ for any intermediate wire in the circuit. Using the $\lambda_\omega$ values for the output wires, the client can translate each of the garbled values returned by the evaluator $p_e$ into a plaintext bit and recover the result of the requested computation. The client declares successful output verification only if the garbled output returned by the evaluator matches with either $\omega_0$ or $\omega_1$, for each output wire $\omega$ of the circuit.}

\underline{\em Collusion Resistance:} Note that, without performing any computation, the evaluator can return one of the two actual garbled outputs for each output wire in the circuit, if and only if it colludes with all the $n$ servers, $\{p_1, p_2, \ldots, p_n\}$, that participated in the creation of the garbled circuit, or factorizes $N$ into its prime factors, $p$ and $q$, which is infeasible.

\underline{\em Unpredictability:} Further, the Blum, Blum, Shub pseudorandom number generator guarantees that one cannot predict the next/previous bit output from the generator, even with the knowledge of all the previous/future bits \cite{blum86,schneier95}. Thus, based on the observations of the garbled values during evaluation, the evaluator cannot predict the preceding or subsequent garbled values, or the $\lambda$ values for any wire in the circuit.

\subsection{Summary of our Proposed System}
\label{sec_our_secure_cloud_computing_model_summary}
We summarize our secure cloud computing model in this subsection.

\begin{enumerate}
    \item{The client chooses a set of $(n+2)$ servers in the cloud, $\{p_1,p_2,\ldots , p_n,p_c,p_e\}$. Then, it provides a description of the desired computation, and a unique seed value $s_i$ to each server $p_i, (1\le i\le n)$. It also provides another seed value $s_{ik}$ to each pair of servers, $(p_i,p_k), (1\le i,k\le n)$.}

    \item{Each server, $p_i, (1\le i\le n)$, creates (or retrieves from its repository, if already available) a Boolean circuit (B) that corresponds to the requested computation.}

    \item{Each server, $p_i, (1\le i\le n)$, uses $s_i$ to generate its shares for the pair of garbled values and a $\lambda$ value for each wire in the circuit ($B$) using the pseudo random generator of Blum, Blum, Shub.

        Using seed $s_i$, each server, $p_i$, generates a pseudorandom bit sequence whose length equals $W(2k+1)$, where $W$ denotes the total number of wires in the Boolean circuit ($B$).}

    \item{The client instructs the $n$ servers, $p_i, (1\le i\le n)$, to use their shares as private inputs for the secure multiparty computation protocol of Goldreich, to construct shares ($GC_i$) of a BMR garbled circuit, $GC$.

        While using the protocol of Goldreich, each pair of servers, $(p_i,p_k), (1\le i,k\le n)$, generates pseudorandom bits using pairwise seed values $s_{ik}$.

        Let $A_i=(A_{00})_i || (A_{01})_i || (A_{10})_i || (A_{11})_i$ denote the shares of server $p_i$ for the four garbled table entries of gate $A$. Then, $GC_i$, in turn, is a concatenation of all bit strings of the form $A_i$, where the concatenation is taken over all the gates in the circuit.}

    \item{The client instructs all $n$ servers, $p_i, (1\le i\le n)$ to send their shares $GC_i$ to the server $p_c$. Performing only XOR operations, the server $p_c$ creates the desired circuit, $GC = GC_1\oplus GC_2\oplus \ldots \oplus GC_n$. Now, the client instructs the server $p_c$ to send the garbled circuit $GC$ to server $p_e$ for evaluation.}

    \item{Using the unique seed values $s_i, (1\le i\le n)$, the client generates garbled input values for each input wire in the circuit, and sends them to the server $p_e$ for evaluation. Using these seed values, the client also generates the $\lambda$ values and the two possible garbled values for each output wire in the circuit, and keeps them secret.}

    \item{Using the garbled inputs, the server $p_e$ evaluates $GC$, and obtains the garbled outputs for each output wire in the circuit and sends them to the client. Using the $\lambda$ values, the client now translates these garbled values into plaintext bits to recover the result of the requested computation.}

    \item{The client checks whether the garbled output for each output wire in the circuit that is returned by the evaluator, $p_e$, matches with one of the two possible garbled values that it computed on its own. If there is a match for all output wires, then the client declares that the evaluator in fact carried out the requested computation.}

\end{enumerate}

\section{Complexity}
\label{sec_complexity}
\subsection{Circuit size of one garbled table entry}

In this section, we analyze the size of the Boolean circuit ($B'$) that computes one specific entry ($A_{ab}$) in the garbled table. Assume that each gate takes two input bits to produce one output bit. Recall from Section~\ref{sec_garbled_circuit_BMR_construction} that $A_{ab} = \gamma_{[((\lambda_x \oplus a) \otimes (\lambda_y \oplus b)) \oplus \lambda_z]} \oplus [G_b(\alpha_{a1})\oplus G_b(\alpha_{a2})\oplus \ldots \oplus G_b(\alpha_{an})]\oplus [G_a(\beta_{b1})\oplus G_a(\beta_{b2})\oplus \ldots \oplus G_a(\beta_{bn})]$, where $\otimes \in \{XOR,AND\}$ denotes the binary operation of gate $A$.

Let $s = ((\lambda_x \oplus a) \otimes (\lambda_y \oplus b)) \oplus \lambda_z$. Computing $s$ requires a total of $3+3(n-1)=3n$ XOR gates and $1$ $\otimes$ gate, since $\lambda_x = \bigoplus_{i = 1}^{n}\lambda_{xi}$, $\lambda_y = \bigoplus_{i = 1}^{n}\lambda_{yi}$, $\lambda_z = \bigoplus_{i = 1}^{n}\lambda_{zi}$, where $\lambda_{xi}, \lambda_{yi}, \lambda_{zi}$ are shares of server $p_i, (1\le i\le n)$.

Boolean circuit $B'$ includes a {\em multiplexer} that chooses $\gamma_s$ using the expression, $\gamma_s=((\gamma_0\oplus \gamma_1).s)\oplus \gamma_0$. This expression is composed of $2$ XOR gates and $1$ AND gate. Since $|\gamma_0|=|\gamma_1|=nk+1$, and  $LSB(\gamma_s)=s$, multiplexing is performed on the most significant $nk$ bits. We can build this multiplexer using a total of $2nk$ XOR gates and $nk$ AND gates.

Now, the expression $\gamma_s\oplus [G_b(\alpha_{a1})\oplus G_b(\alpha_{a2})\oplus \ldots \oplus G_b(\alpha_{an})]\oplus [G_a(\beta_{b1})\oplus G_a(\beta_{b2})\oplus \ldots \oplus G_a(\beta_{bn})]$ has $(2n+1)$ terms, which are combined using $2n$ XOR operations. Since, each term has a length of $(nk+1)$ bits, computing this expression requires a total of $2n(nk+1)$ XOR gates.

To summarize, the Boolean circuit ($B'$) that computes one specific garbled table entry ($A_{ab}$) has a total of $(3n+2nk+2n(nk+1))$ XOR gates, $nk$ AND gates, and $1$ $\otimes$ gate.

\begin{figure}[t]
\centering
\includegraphics[width=3.25in]{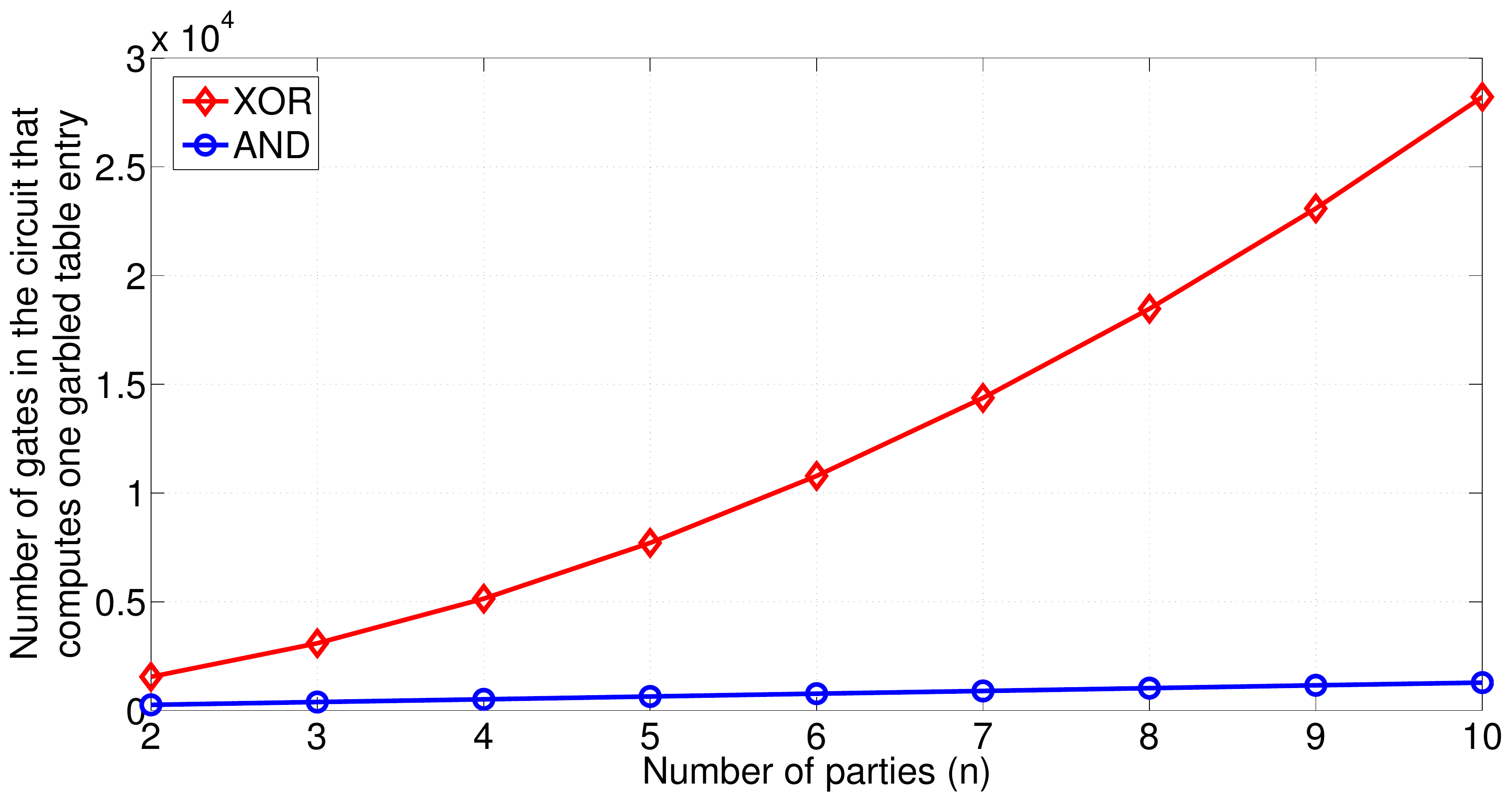}
\vspace{-3 mm}
\caption{Number of gates in the circuit that computes one garbled table entry as a function of $n$.\vspace{-4 mm}}
\label{fig_num_gates_aux_bool_circuit}
\end{figure}

Fig.~\ref{fig_num_gates_aux_bool_circuit} shows the total number of gates in the circuit that computes $A_{ab}$, when $A$ is an AND gate, as a function of $n$ for a fixed value of $k=128$ bits. Notice the relatively small number of AND gates in the circuit. For example, when $n=6$, the circuit that computes $A_{ab}$ has a total of $10782$ XOR and $769$ AND gates. {\em While the number of XOR gates increases quadratically with $n$, the number of AND gates increases only linearly with $n$}.

Let $B$ denote the Boolean circuit that corresponds to the desired computation such as addition of two numbers. Then, while creating the garbled circuit $GC$ for $B$, the $n$ parties use the circuit $B'$ for the protocol of Goldreich to compute each one of the four garbled table entries of the form $A_{ab}$ for each gate $A$ in the circuit $B$.



\subsection{Communication cost to compute one garbled table entry}
\label{sec_communication_cost_to_compute_one_entry}

A 1-out-of-2 OT exchange between two parties involves the exchange of: (i) a random element $C$ from the prime order subgroup, $G$, of $Z_p^*$; (ii) a public key, $PK_0$; and (iii) the encryptions, $E_0,E_1$, of the plaintext messages $M_0,M_1$. Let $k$ denote the security parameter, which equals the size of the plaintext messages, $M_0,M_1$. Let $s_{1:2}$ denote the total number of bits that are exchanged during a 1-out-of-2 OT. Then, $s_{1:2}=|C|+|PK_0|+|E_0|+|E_1|=|p|+|p|+(|p|+k)+(|p|+k)=4|p|+2k$.

A 1-out-of-4 OT exchange between two parties includes: (i) two 1-out-of-2 OTs, and (ii) four encryptions, $E_{00},E_{01},E_{10},E_{11}$. Let $s_{1:4}$ denote the total number of bits that are exchanged during a 1-out-of-4 OT. Then, $s_{1:4}=2(s_{1:2})+4k=8(|p|+k)$. Note that $|p|$ and $k$ are public and symmetric key security parameters, respectively. For example, $|p|=3072$ achieves the equivalent of $k=128$-bit security \cite{nist_key_len_recommendations2012}; in this case, the sum of the sizes of all messages exchanged during a 1-out-of-4 OT is $s_{1:4}=3200$ bytes.

For each AND gate in the circuit $B'$, all possible pairs of the $n$ servers, $(p_i, p_j), 1\le i<j\le n$, engage in a 1-out-of-4 OT, and there are a total of $n(n-1)/2$ combinations of $(p_i,p_j)$. Since the number of AND gates in the circuit $B'$ is at most $(nk+1)$, the total number of 1-out-of-4 OTs is $t_{1:4}=(nk+1)\times n(n-1)/2$.

At the completion of the secure multiparty computation protocol of Goldreich, each server, $p_i, (1\le i\le n)$, sends its share $(A_{ab})_i$ to another server, $p_c$, to create the desired garbled table entry, $A_{ab}$. Since $|(A_{ab})_i|=nk+1$, the server $p_c$ receives a total of $s^{\star}=n(nk+1)$ bits from the other $n$ servers.

To summarize, in order to create one entry, $A_{ab}$, the total amount of network traffic, $T=(t_{1:4}\times s_{1:4})+s^{\star}=(nk+1)[(4(|p|+k)\times n(n-1))+n]$. {\em When the security parameters, $k$ and $|p|$, are fixed, the network traffic is a cubic function of $n$.}

\begin{figure}[t]
\centering
\includegraphics[width=3.25in]{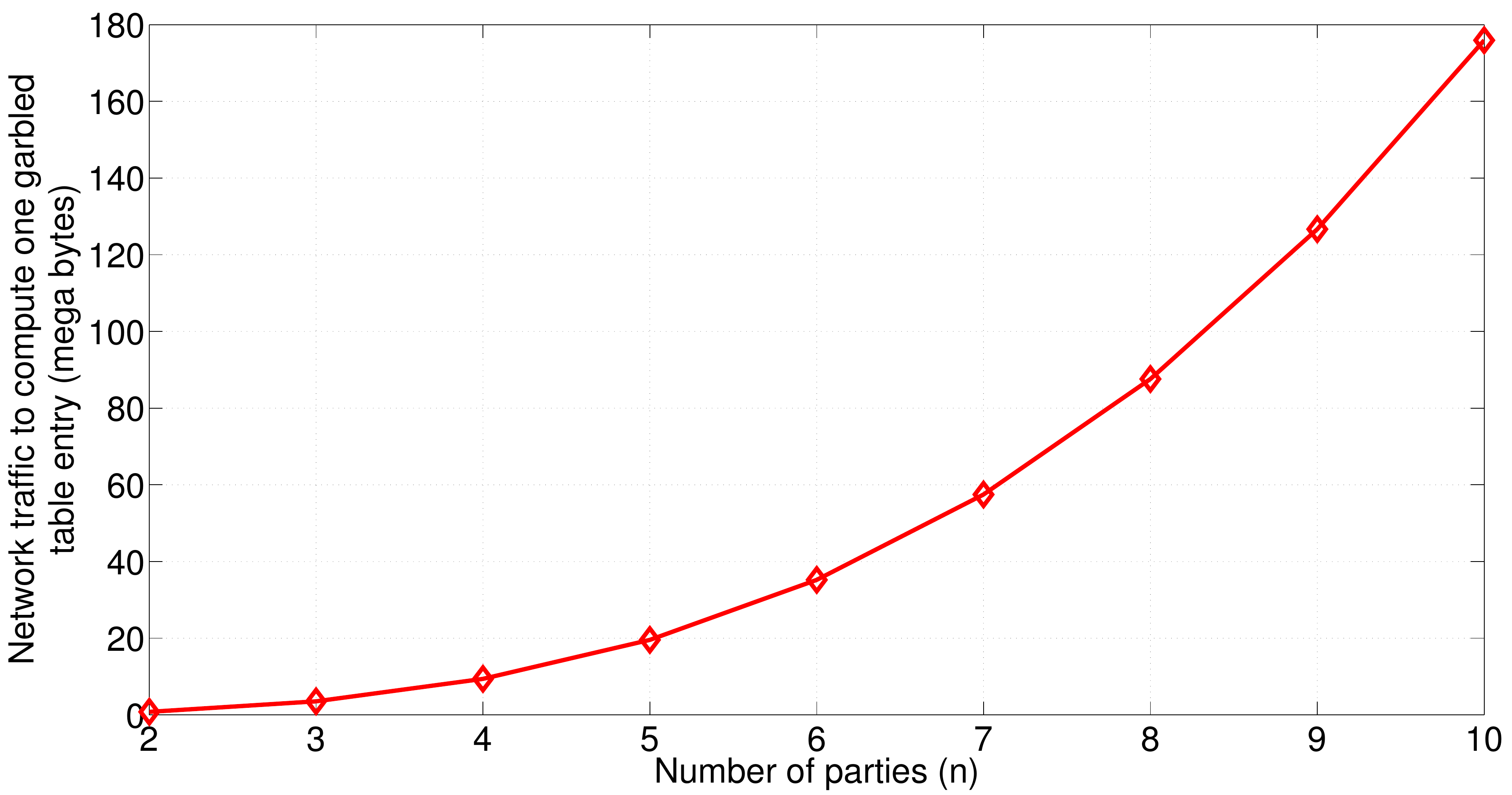}
\vspace{-3 mm}
\caption{Amount of network traffic to compute one garbled table entry as a function of $n$.\vspace{-4 mm}}
\label{fig_network_traffic}
\end{figure}

Fig.~\ref{fig_network_traffic} shows the network traffic to create one garbled table entry as a function of $n$. For example, when $n=5$, the cloud servers exchange a total of $19.56$ MB of data in order to create a single entry in the garbled table.

Let $N_g$ denote the total number of gates in the circuit $B$ that corresponds to the desired computation. Then, in the process of creating the garbled circuit, $GC$, the total amount of network traffic equals $4N_g\times T$.

\subsection{Computation cost of creating the garbled circuit}

Let $W$ denote the total number of wires in the circuit $B$. For each wire, each server, $p_i, (1\le i\le n)$, generates $(2k+1)$ bits using the Blum, Blum, Shub (BBS) pseudorandom number generator (PRNG) for its share of garbled values and the $\lambda$ value. Therefore, the $n$ servers collectively generate a total of $b_{1}=n(2k+1)W$ bits using the BBS PRNG. Let $N$ denote the modulus value in BBS PRNG (note: $|N|=3072$ achieves $128$-bit security \cite{nist_key_len_recommendations2012}). Then, $n(2k+1)W$ modular multiplication operations in $Z_N^*$ are necessary to generate bits using BBS PRNG.

Let $W_o$ denote the number of output wires in the circuit $B$. Let $G$ denote the PRNG, which we have described in Section~\ref{sec_garbled_circuit_BMR_evaluation}, that outputs a sequence of $(2nk+2)$ bits on providing a $k$-bit input seed. Each server uses the PRNG $G$, on its share of each garbled value for every non-output wire in the circuit $B$. Then, in the process of creating the garbled circuit, the $n$ servers collectively use the PRNG $G$, $2n(W-W_o)$ times to generate a total of $b_{2}=4n(nk+1)(W-W_o)$ bits.

Let $N_g$ denote the total number of gates in the circuit $B$. For protocol of Goldreich, the total number of pseudorandom bits generated by each party using the PRNG $R$, equals $8(n-1)m\times N_g$, where $m=(3+2(nk+1)+2k)$ (Section~\ref{sec_key_changes_to_BMR_Goldreich}). Thus, the $n$ parties collectively generate a total of $b_{3}=8n(n-1)(3+2(nk+1)+2k)\times N_g$ bits using the PRNG $R$.

Note that both the PRNG $G$ and PRNG $R$ can be realized using a block cipher such as AES operating in output feedback mode.

Let $t_{1:4}$ denote the number of 1-out-of-4 OTs to create one garbled table entry (Section~\ref{sec_communication_cost_to_compute_one_entry}). Then, the total number of 1-out-of-4 OTs to create the complete garbled circuit $GC$ is at most $4N_g\times t_{1:4}=4N_g\times (nk+1)\times n(n-1)/2$.

During a 1-out-of-2 oblivious transfer, the sender and chooser generate a total of $|C|+|k|+|r_0|+|r_1|=(4|p|-1)$ bits. Each 1-out-of-4 oblivious transfer involves the cost of two 1-out-of-2 oblivious transfers, in addition to generating a total of $|L_0|+|L_1|+|R_0|+|R_1|=4k$ bits. {\em A very small constant number of modular arithmetic operations, AES and SHA crypto operations are carried out during each OT} (Section~\ref{sec_1_out_of_2_OT} and Section~\ref{sec_1_out_of_4_OT}). While creating the garbled circuit $GC$, these sets of operations are performed $4N_g\times (nk+1)\times n(n-1)/2$ times; and a total of $b_{4}=4N_g\times (nk+1)\times (n(n-1)/2)\times (8|p|+4k-2)$ bits are generated during the OTs.



To summarize, the total number of bits that are generated by the $n$ parties while creating the garbled circuit is $b=b_1+b_2+b_3+b_4=(n(2k+1)W)+(4n(nk+1)(W-W_o))+(8n(n-1)(3+2(nk+1)+2k)\times N_g)+(4N_g\times (nk+1)\times (n(n-1)/2)\times (8|p|+4k-2))$. {\em Thus, for any given Boolean circuit, when the security parameters $k$ and $|p|$ are fixed, the total number of bits generated randomly is a cubic function of $n$.}

\begin{figure}[t]
\centering
\includegraphics[width=3.25in]{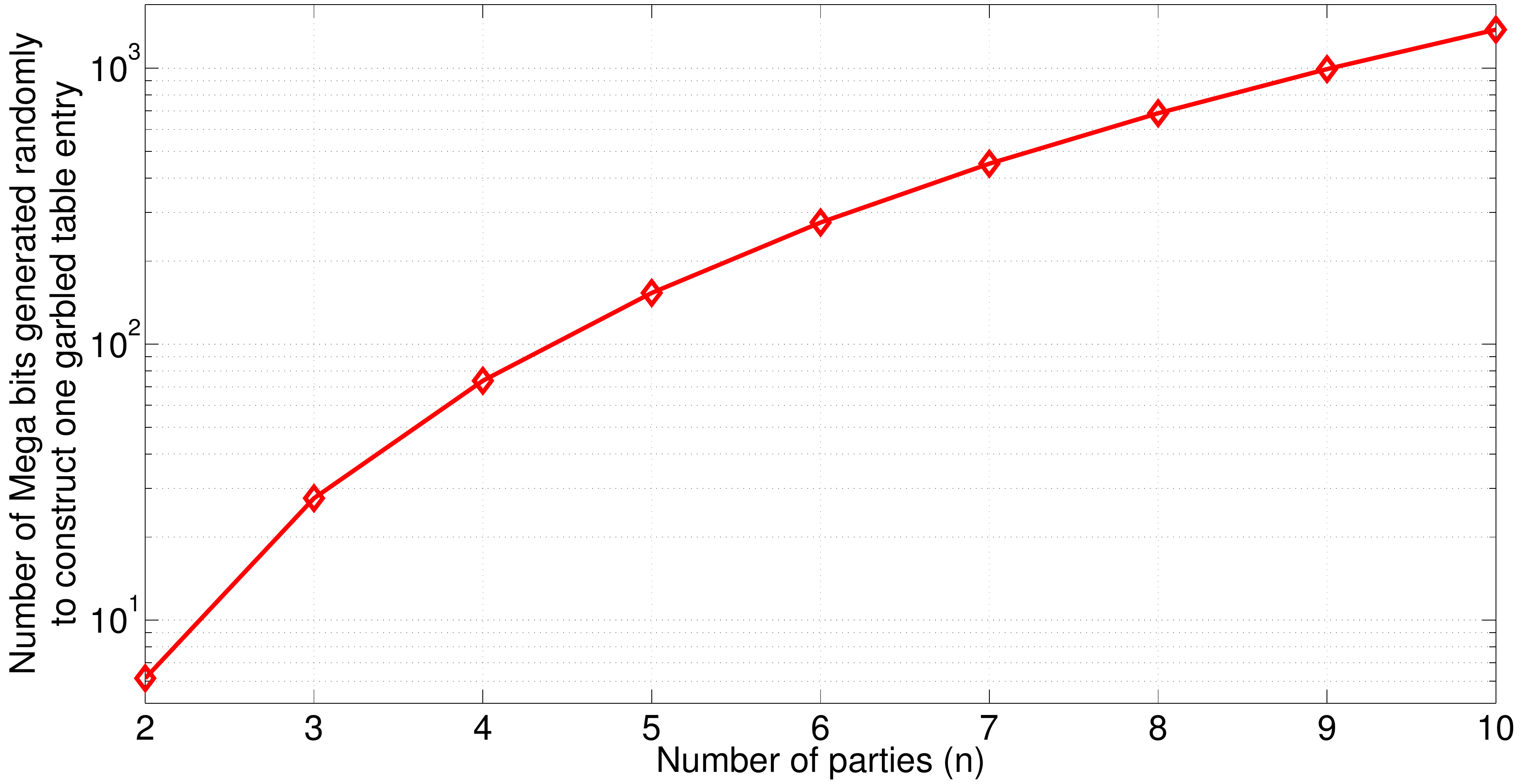}
\vspace{-3 mm}
\caption{Total number of Mega bits that are generated randomly in the process of creating one garbled table entry for the $32$-bit adder.\vspace{-2 mm}}
\label{fig_num_rand_bits}
\end{figure}

For example, consider the construction of a garbled circuit for adding two $32$-bit numbers. The corresponding Boolean circuit has a total of $W=439$ wires, $W_o=33$ output wires, and $N_g=375$ gates\footnote{The $32$-bit adder circuit in \cite{boolean_circuits} has $127$ AND gates, $61$ XOR gates and $187$ NOT gates. Note that a NOT gate is equivalent to an XOR gate, since NOT$(x)=(1\oplus x$).}. Fig.~\ref{fig_num_rand_bits} shows the total number of Mbits that are generated randomly while creating one garbled table entry (i.e., $b/(4N_g\times 2^{20})$) for the $32$-bit adder. As an example, when $n=5$, these parties collectively generate a total of $153.41$ Mbits, randomly, to create one garbled table entry.

\subsection{Cost of evaluating the garbled circuit}

In order to perform the requested computation, the server $p_e$ obtains the garbled circuit, $GC$, from the server $p_c$. Let $N_g$ denote the total number of gates in the circuit $B$. Each entry in the garbled table has a length of $(nk+1)$ bits. Therefore, the size of the garbled circuit equals $4N_g\times (nk+1)$ bits.

\begin{figure}[t]
\centering
\includegraphics[width=3.25in]{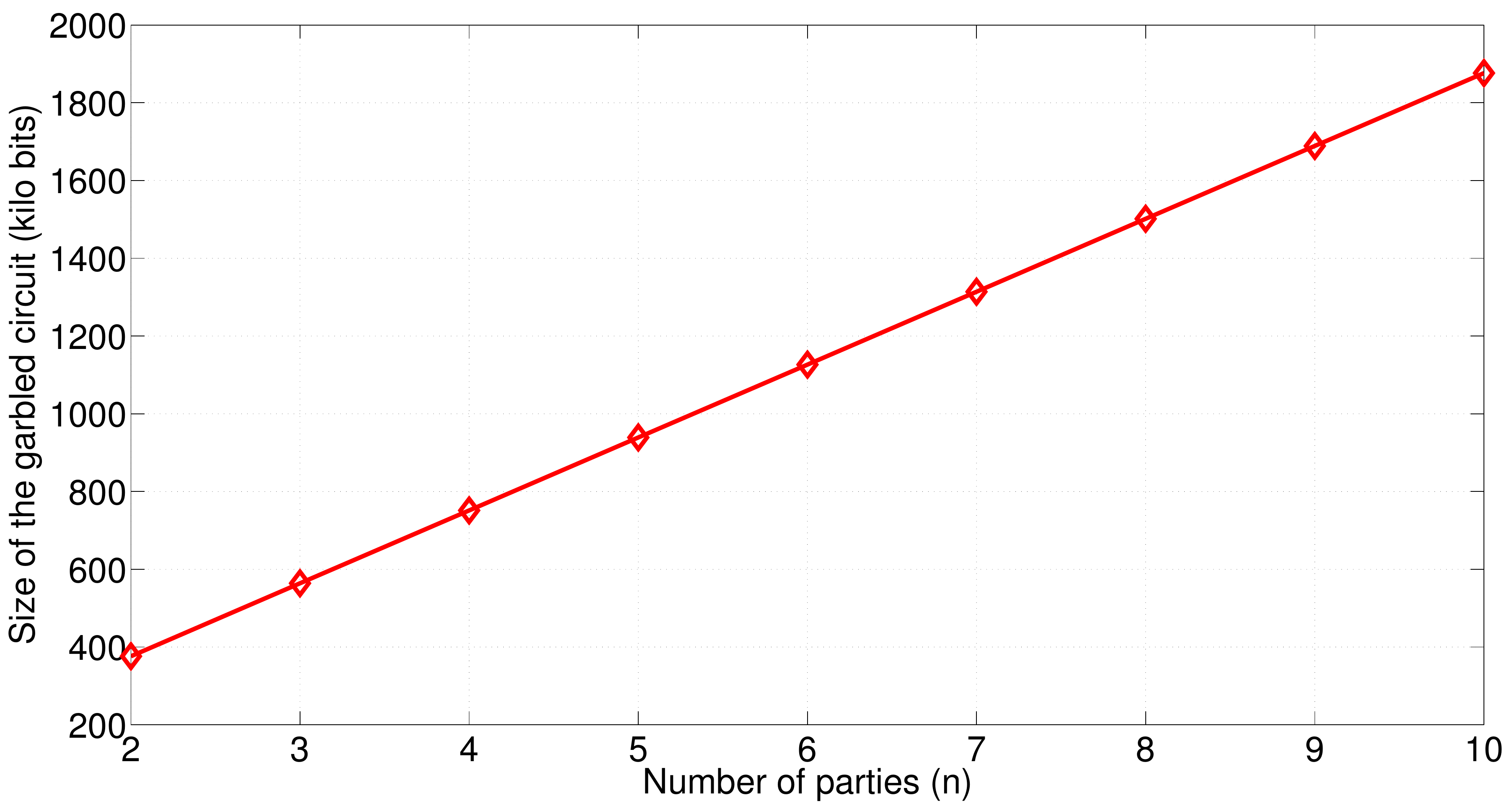}
\vspace{-3 mm}
\caption{Size of the garbled circuit in kilo bits for the $32$-bit adder.\vspace{-3 mm}}
\label{fig_size_of_garbled_circuit}
\end{figure}

Fig.~\ref{fig_size_of_garbled_circuit} shows the size of the garbled circuit in kilo bits for the 32-bit adder. This circuit has $N_g=375$ gates. The security parameter $k=128$.

Let $W$ and $W_o$ denote the total number of wires and output wires, respectively, in the Boolean circuit $B$. During evaluation, for each non-output wire of the circuit, the server $p_e$ uses the PRNG $G$ $n$ times. Therefore, $G$ is used for a total of $(W-W_o)n$ times.

\subsection{Cost for the client}
\label{sec_cost_for_client}
To enable the creation of the garbled circuit, the client provides: (i) a unique seed value, $s_i$, to each server $p_i, (1\le i\le n)$, and (ii) a seed value, $s_{ik}$, to each pair of servers $(p_i,p_k),(1\le i,k\le n)$.

For the BBS PRNG, the length of each seed value, $|s_i|=|N|$. For the PRNG $R$, which can be implemented using a block cipher such as AES in output feedback mode, the length of each seed is $|s_{ik}|=k$. Therefore, the total number of bits that the client exchanges for the seed values is $b_s=n|N|+n(n-1)k=n(|N|+(n-1)k)$.

For each plaintext input bit to the circuit, the client is required to generate the garbled input. Each garbled value is $(nk+1)$ bits long, whose least significant bit depends on the $\lambda$ value. Since the $\lambda$ value, in turn, depends on the $1$-bit shares for the $n$ parties, the number of bits that the client needs to generate for each input wire equals $b_i=(nk+n)$.

To enable verification of outputs, the client needs to generate both possible garbled outputs for each output wire. Therefore, the number of bits that the client needs to generate for each output wire equals $b_o=(2nk+n)$.

To summarize, the client generates/exchanges a total of $b_s+W_i\times b_i+W_o\times b_o=n[|N|+(n-1)k+W_i(k+1)+W_o(2k+1)]$ bits, where $W_i$ and $W_o$ denote the number of input and output wires, respectively, in the Boolean circuit $B$.

\begin{figure}[t]
\centering
\includegraphics[width=3.25in]{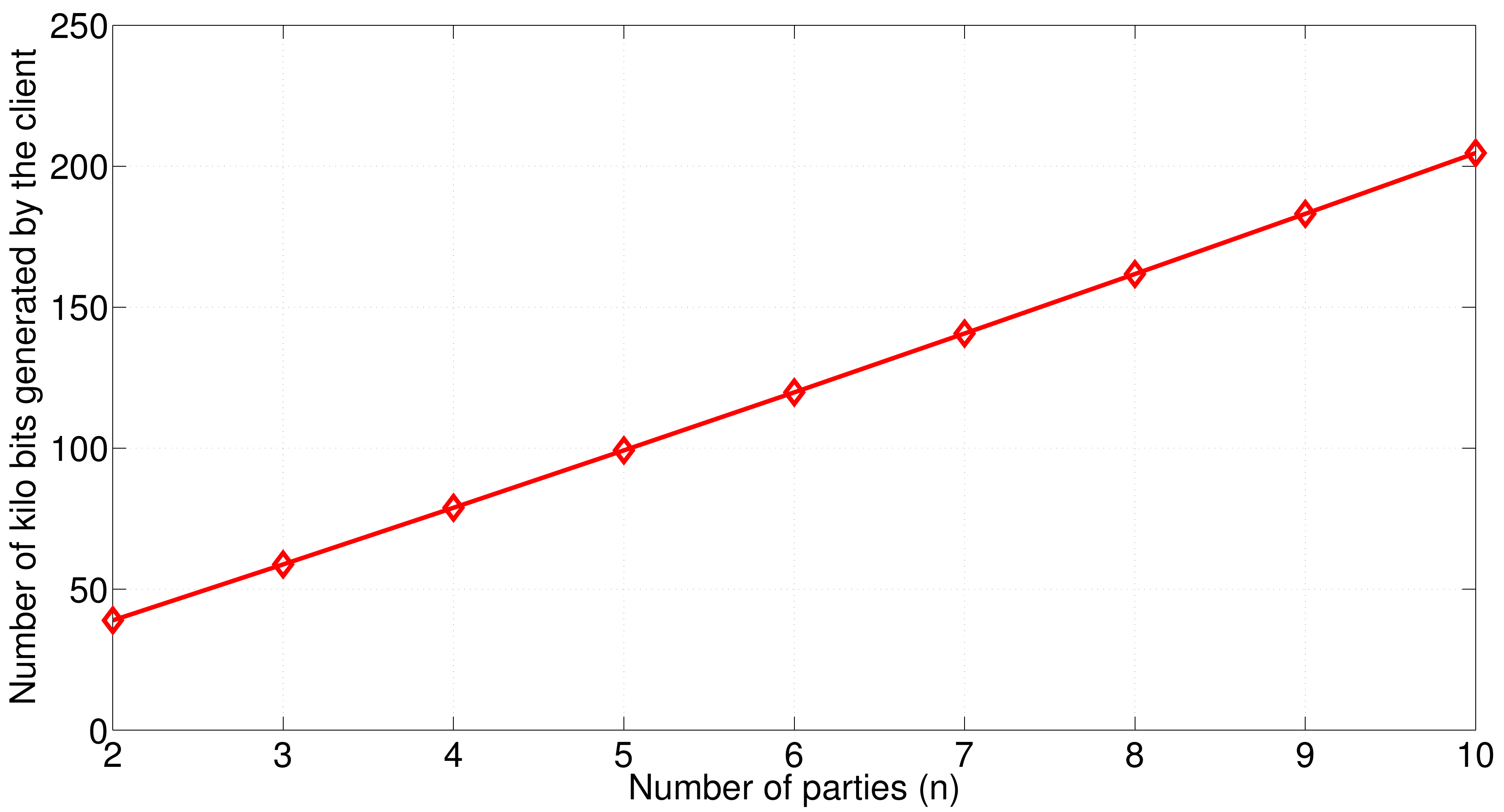}
\vspace{-3 mm}
\caption{Total number of kilo bits that the client generates to delegate the construction and evaluation of the garbled circuit, and to verify the outputs for the $32$-bit adder.\vspace{-3 mm}}
\label{fig_num_bits_client}
\end{figure}

Fig.~\ref{fig_num_bits_client} shows the total number of kilo bits that the client generates in order to enable the servers to construct and evaluate the garbled circuit, as well as for the verification of outputs for the $32$-bit adder. This circuit has $W_i=64$ input wires and $W_o=33$ output wires. The security parameters are $k=128$ and $|N|=3072$.

Comparing Fig.~\ref{fig_num_bits_client} with Figs.~\ref{fig_network_traffic} and Fig.~\ref{fig_num_rand_bits}, we notice that while the servers generate and exchange Gigabytes of information to create the garbled circuit, the mobile client, on the other hand, generates and exchanges only kilobytes of information with the evaluator and the other servers.

\subsection{Comparison of Our Scheme with Gentry's FHE Scheme}
While Gentry's FHE scheme \cite{gentry10} uses only one server, it, however, requires the client to exchange $O(k^5)$ bits with the evaluating server, for each input and output wire of the circuit. In our secure cloud computing system, since each garbled value has a length of $(nk+1)$ bits, for each input and output wire, the client only exchanges $O(nk)$ bits with the server $p_e$. For example, with $k=128$, the size of each encrypted plain text bit equals several Gigabits with Gentry's scheme, while it equals a mere $641$ bits in our approach with $n=5$. {\em Thus, our approach is far more practical for cloud computing in mobile systems in comparison to FHE schemes.}

\subsection{Construction and evaluation time}

\begin{figure}[t]
\centering
\includegraphics[width=3.25in]{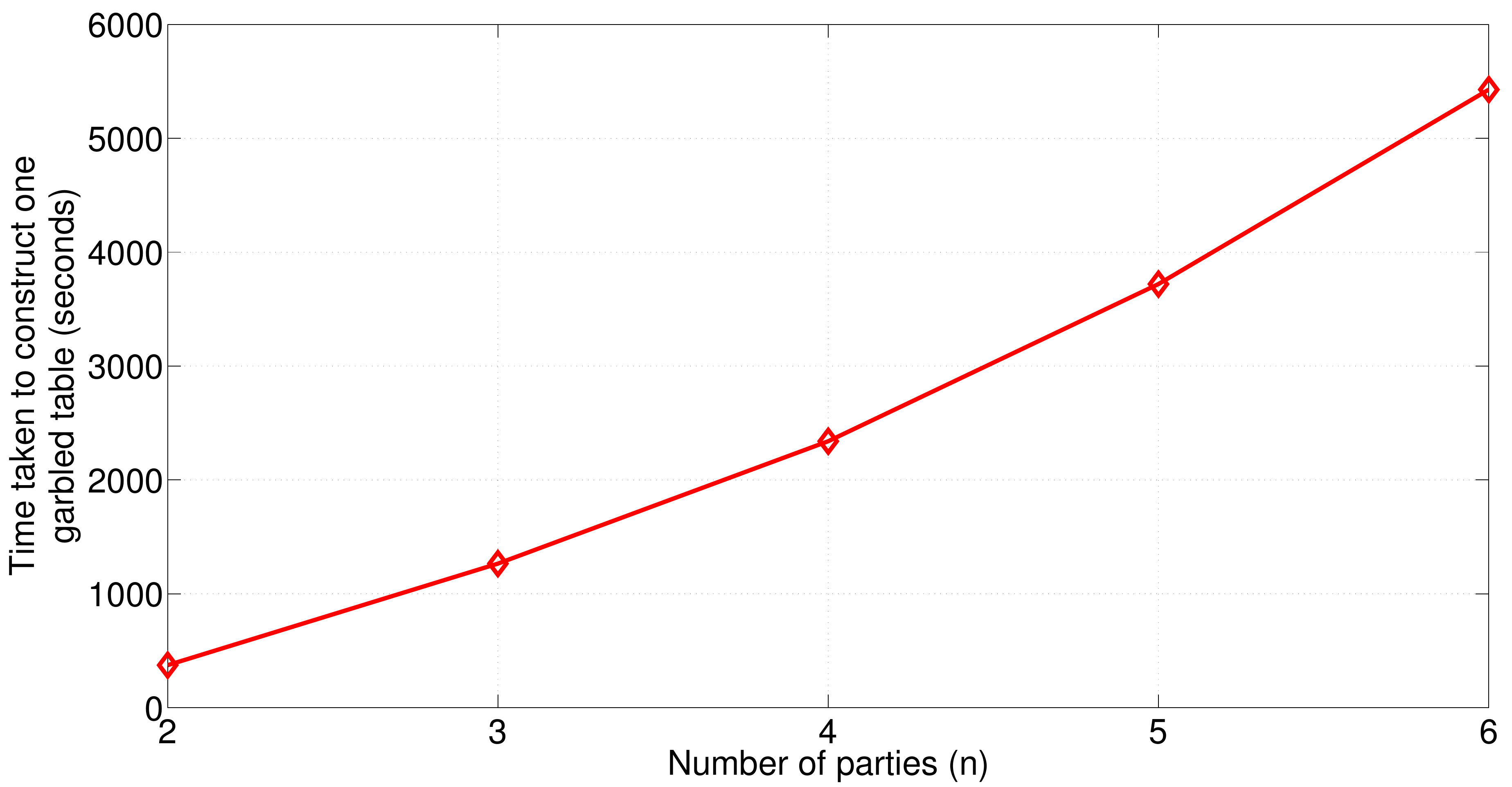}
\vspace{-3 mm}
\caption{Time taken to construct one garbled table.\vspace{-2 mm}}
\label{fig_time_construction}
\end{figure}

We implemented our secure cloud computing system using BIGNUM routines and crypto functions from the OpenSSL library. We built our system as a collection of modules, and the servers communicate using TCP sockets. We evaluated our system on a server with Intel Xeon $2.53$ GHz processor, with $6$ cores and $32$ GB RAM \footnote{We thank Srinivasa Vamsi Laxman Perala (email: sxp136630@utdallas.edu) for implementing various components of our system.}. Fig.~\ref{fig_time_construction} shows the time taken to construct one garbled table as a function of $n$. {\em We note that the garbled tables for any number of gates in the circuit can be constructed in parallel, which will significantly reduce the construction time.}

\begin{figure}[t]
\centering
\includegraphics[width=3.25in]{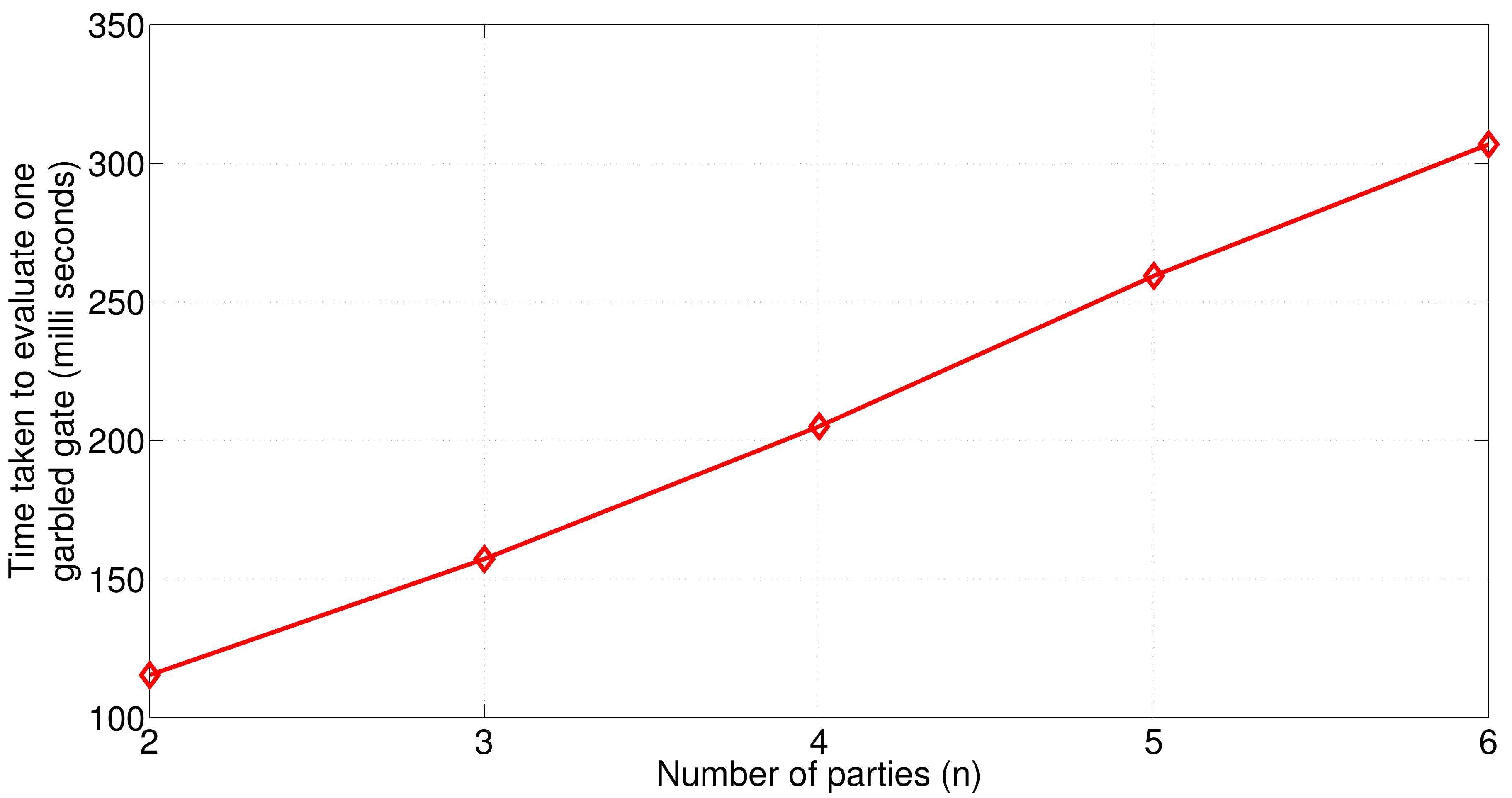}
\vspace{-3 mm}
\caption{Time to evaluate one garbled gate.\vspace{-3 mm}}
\label{fig_time_evaluation}
\end{figure}

Fig.~\ref{fig_time_evaluation} shows the time taken to evaluate one garbled gate as a function of $n$. Comparing Fig.~\ref{fig_time_construction} and Fig.~\ref{fig_time_evaluation}, we observe that {\em evaluation is significantly faster than construction, where the latter can be done offline.} If the garbled circuits are {\em pre-computed}, and made available to the evaluator, in advance, it can readily carry out the requested computation, and therefore, drastically reduce the {\em response time} for the mobile client.

\section{Privacy Preserving Search for the Nearest Bank/ATM}
\label{sec_manhattan_distance}

We examine the following privacy preserving application in this section. A mobile client, which is located at the intersection of two streets, needs to determine the location of the nearest Chase or Wells Fargo bank or ATM machine in a privacy preserving manner. We evaluate our application using real-world data available for Salt Lake City, UT, whose streets are arranged in a {\em grid pattern}. Our application assures the privacy of the following -- (i) the mobile client's input location, (ii) the computed bank/ATM location nearest to the client, and (iii) the computed distance to the nearest ATM. Note that these secrets are revealed to the evaluator only if it colludes with all the $n$ servers that participate in the creation of the garbled circuit.

\begin{table}[!t]
\renewcommand{\arraystretch}{1.3}
\caption{Locations of Banks and ATMs in Salt Lake City, UT {\em (source: www.chase.com, www.wellsfargo.com).}}
\label{table_atm_locations}
\centering
\begin{tabular}{||c||c||}
\hline
\bfseries Bank/ATM & \bfseries Location\\
\hline
Chase & $201$ South $0$ East\\
\hline
Chase & $185$ South $100$ East\\
\hline
Chase & $376$ East $400$ South\\
\hline
Chase & $531$ East $400$ South\\
\hline
Wells Fargo & $299$ South $0$ East\\
\hline
Wells Fargo & $381$ East $300$ South\\
\hline
Wells Fargo & $79$ South $0$ East\\
\hline
Wells Fargo & $778$ South $0$ East\\
\hline
Wells Fargo & $570$ South $700$ East\\
\hline
Wells Fargo & $235$ South $1300$ East\\
\hline
\end{tabular}
\end{table}

We consider an area of Salt Lake City, UT that lies between Main street (which represents $0$ East street), $1300$ East street, South Temple street (which represents $0$ South street), and $800$ South street. This area consists of $L=10$ ATM locations that are shown in Table~\ref{table_atm_locations}.

Each East/South coordinate in this area is an $l=max(\lceil log_2 1300\rceil, \lceil log_2 800\rceil) = 11$-bit unsigned number. Therefore, the location of the mobile client at an intersection, or any bank/ATM in this area can be identified using $L_{ind}=2l=22$ bits.

\subsection{Circuit for Computing Manhattan Distance}

Let $(x_a, y_a)$ represent the coordinates of the mobile client at an intersection. Similarly, let $(x_b, y_b)$ represent the coordinates of a bank/ATM. Since the streets are arranged in a grid pattern, the shortest distance ($D$) between $(x_a, y_a)$ and $(x_b, y_b)$ equals the sum of the absolute differences between the respective coordinates: $D=|x_a - x_b|+|y_a - y_b|$. This distance metric is more commonly referred to as the Manhattan distance.

\begin{figure}[t]
\centering
\includegraphics[width=3.35in]{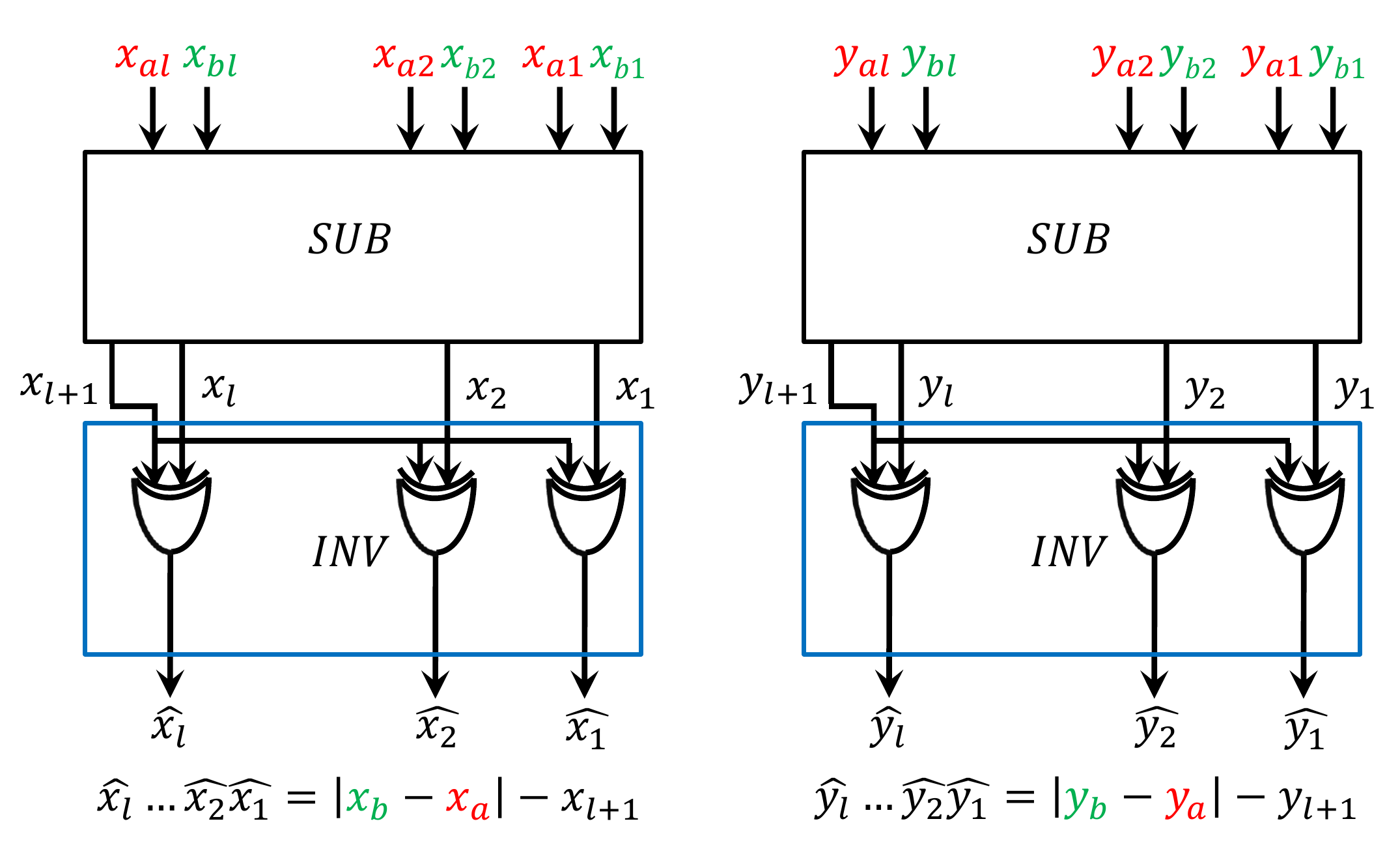}
\caption{Absolute difference between the $X$ and $Y$ coordinates.}
\label{fig_absolute_diff}
\end{figure}

\begin{figure}[t]
\centering
\includegraphics[width=3.0in]{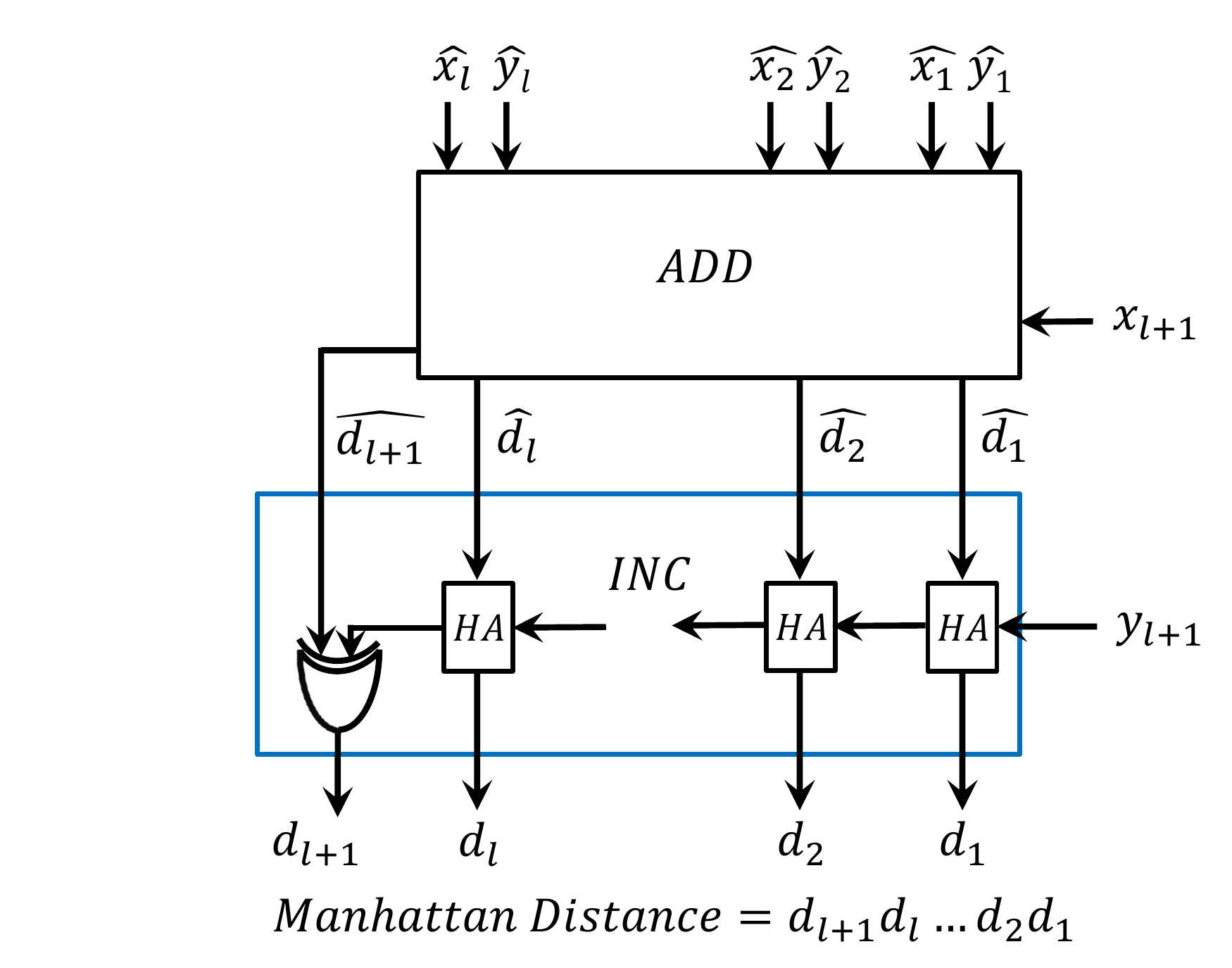}
\caption{Manhattan Distance Computation.}
\label{fig_manhattan_dist_final_step}
\end{figure}

We design a Boolean circuit for computing the Manhattan distance between two points, $(x_a, y_a)$ and $(x_b, y_b)$. Assume that each coordinate is an $l$ bit unsigned number. Figure~\ref{fig_absolute_diff} and Figure~\ref{fig_manhattan_dist_final_step} show the block diagram of our circuit. We use the $SUB$ and $ADD$ blocks of Kolesnikov et al. \cite{kolesnikov09}.

Each $SUB$/$ADD$ block is composed of $l$ $1$-bit subtractors/adders; each $1$-bit subtractor/adder, in turn, is composed of $4$ XOR gates, $1$ AND gate. Note that $(l+1)^{th}$ output bit of $SUB$ block equals the complement of carry-out bit from the $l^{th}$ $1$-bit subtractor.

If $x_a \geq x_b$, then the output of $SUB$ block equals the absolute difference, $|x_a - x_b|$. Otherwise, the output of $SUB$ block equals the negative value, $-|x_a - x_b|$, in $2's$ complement form. Since $x_{l+1}=1$ for negative values, we use $x_{l+1}$ as one of the inputs for the $XOR$ gates in the $INV$ block to compute the $1's$ complement of the absolute difference. Then, we subsequently use $x_{l+1}$ as a carry-in input bit for the $ADD$ block. Thus, the output of the $ADD$ block accounts for both $x_a \geq x_b$ and $x_a < x_b$ cases.

Similarly, for the $Y$ coordinates we use the $SUB$ and $INV$ blocks to compute the absolute difference in $1's$ complement form. To account for the case when $y_{l+1}=1$, we use an $INC$ block that adds the value of the bit $y_{l+1}$ to the output of the $ADD$ block. The $INC$ block is composed of $l$ half-adder(HA) blocks, where each $HA$ block, in turn, is composed of $1$ $XOR$ and $1$ $AND$ gate.

\begin{table}[!t]
\renewcommand{\arraystretch}{1.3}
\caption{Circuit size for Manhattan distance calculation.}
\label{table_manhattan_dist_calc}
\centering
\begin{tabular}{||c||c||c||}
\hline
\bfseries Block & \bfseries $\#XOR$ gates & \bfseries $\#AND$ gates\\
\hline
$2\times SUB$ & $2\times 4l$ & $2\times l$\\
\hline
$2\times INV$ & $2\times l$ & $0$\\
\hline
$ADD$ & $4l$ & $l$\\
\hline
$INC$ & $l+1$ & $l$\\
\hline
$Total$ & $15l+1$ & $4l$\\
\hline
\end{tabular}
\end{table}

The output of the $INC$ block represents the $(l+1)$-bit Manhattan distance between $(x_a,y_a)$ and $(x_b,y_b)$. Using the circuit design of Figure~\ref{fig_absolute_diff} and Figure~\ref{fig_manhattan_dist_final_step}, a list of distances between the location of the mobile client and any number ($L$) of ATM locations can be computed. Table~\ref{table_manhattan_dist_calc} shows that our Boolean circuit for computing the Manhattan distance between two points has a total of $(15l+1)$ $XOR$ gates and $4l$ $AND$ gates.

{\em Note:} In a more direct alternative to compute Manhattan distance, we may first find $min(x_a,x_b)$ and $max(x_a,x_b)$, and always subtract $min(x_a,x_b)$ from $max(x_a,x_b)$ (similarly for the $Y$ coordinates). While this approach would eliminate $INV$ and $INC$ blocks, it however would require the use of two comparator and conditional swap blocks \cite{huang2012private}, which together introduce $12l$ new $XOR$ gates and $4l$ $AND$ gates. Consequently, this alternative approach to compute Manhattan distance would require a total of $24l$ $XOR$ gates and $7l$ $AND$ gates. Thus, in comparison to this more direct alternative, our design shown above in Figure~\ref{fig_absolute_diff} and Figure~\ref{fig_manhattan_dist_final_step} requires a significantly smaller number of $(15l+1)$ $XOR$ and $4l$ $AND$ gates.

\subsection{Circuit for Computing the Nearest ATM}

\begin{figure}[t]
\centering
\includegraphics[width=3.0in]{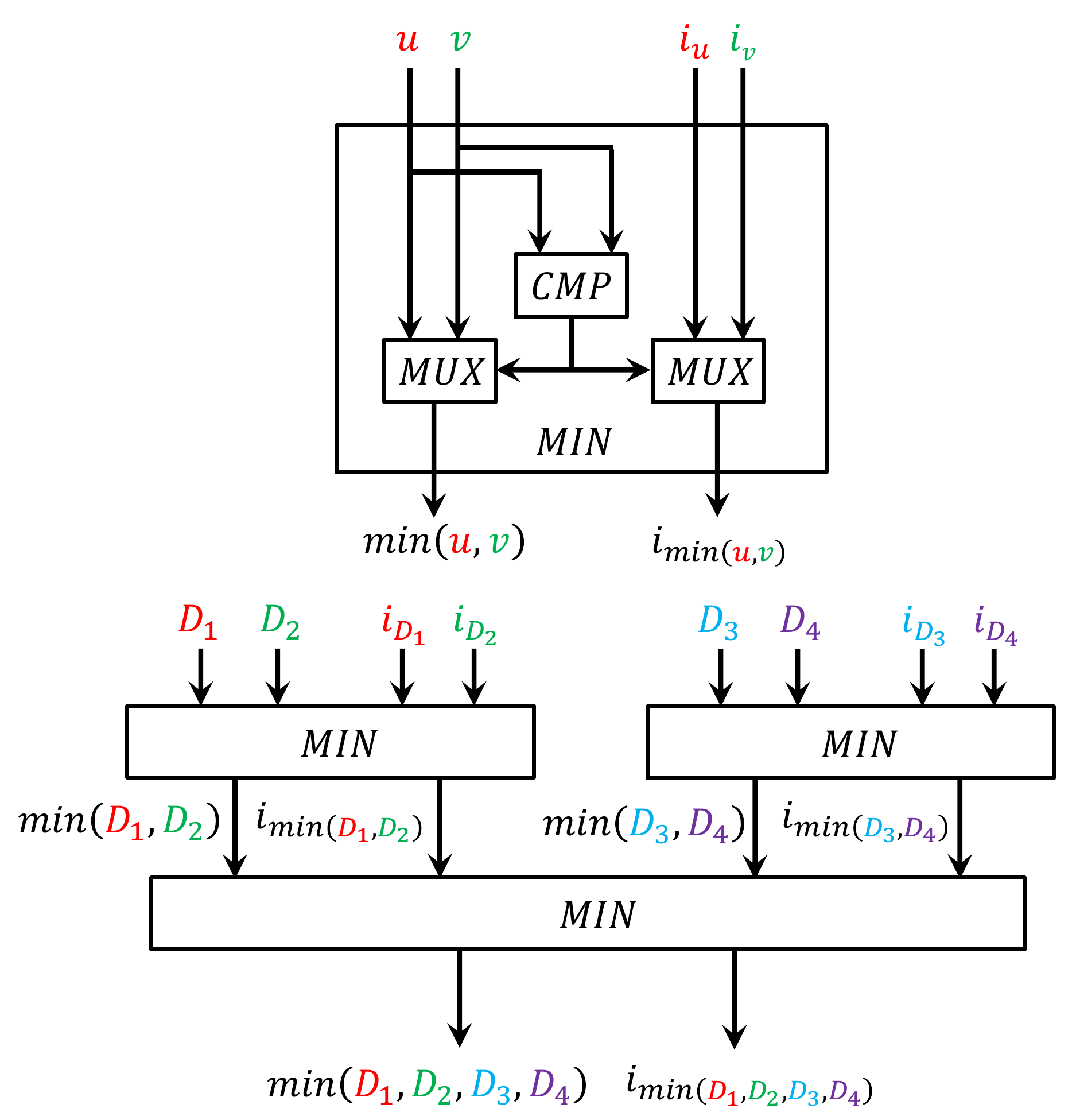}
\caption{Finding minimum value and index.}
\label{fig_min_tree}
\end{figure}

\begin{table}[!t]
\renewcommand{\arraystretch}{1.3}
\caption{Circuit size of $1$ Min block.}
\label{table_one_min_value_index_block}
\centering
\begin{tabular}{||c||c||c||}
\hline
\bfseries Block & \bfseries $\#XOR$ gates & \bfseries $\#AND$ gates\\
\hline
$CMP$ & $3(l+1)$ & $l+1$\\
\hline
$MUX_{min}$ & $2(l+1)$ & $l+1$\\
\hline
$MUX_{index}$ & $2L_{ind}$ & $L_{ind}$\\
\hline
$Total$ & $5(l+1)+2L_{ind}$ & $2(l+1)+L_{ind}$\\
\hline
\end{tabular}
\end{table}

Following the computation of distance between the mobile client and $L$ ATM locations, it is necessary to find the nearest ATM, along with its distance. We use the approach of Kolesnikov et al. \cite{kolesnikov09} to find the minimum value and its index, given a list of values. Kolesnikov et al. have designed a $MIN$ block to find the minimum of two input values -- it uses the result of a comparator to multiplex the minimum value, as well as the corresponding index as shown in Figure~\ref{fig_min_tree}. Table~\ref{table_one_min_value_index_block} shows the size of the circuit that computes the minimum of two values, along with its index. In our privacy preserving application, each distance is an $(l+1)=11+1=12$-bit number, and each index that identifies an ATM using its East and South coordinates is an $L_{ind}=2l=22$-bit number.

Given $L$ values and their indices, using $(L-1)$ $MIN$ blocks organized as a tree, the minimum value and the corresponding index are propagated from the leaves to the root. Figure~\ref{fig_min_tree} shows the computation of the minimum of $4$ input values, $D_1$, $D_2$, $D_3$, $D_4$, and the corresponding index, $i_{min(D_1,D_2,D_3,D_4)}$, as an example.

\begin{table}[!t]
\renewcommand{\arraystretch}{1.3}
\caption{Complete circuit size for nearest ATM. $DIST(L)$ denotes distance to $L$ locations. $T(MIN)$ denotes the tree of MIN blocks.}
\label{table_complete_circuit_closest_ATM}
\centering
\begin{tabular}{||c||c||c||}
\hline
\bfseries Block & \bfseries $\#XOR$ gates & \bfseries $\#AND$ gates\\
\hline
$DIST(L)$ & $v_1=(15l+1)L$ & $v_3=4lL$\\
\hline
$T(MIN)$ & $v_2=[5(l+1)+2$ & $v_4=[2(l+1)+$\\
 & $L_{ind}]\times (L-1)$ & $L_{ind}]\times (L-1)$\\
\hline
$Total$ & $v_1+v_2$ &$v_3+v_4$\\
\hline
\hline
$Application$ & $2596$ & $854$\\
\hline
\end{tabular}
\end{table}

Table~\ref{table_complete_circuit_closest_ATM} shows the number of $XOR$ and $AND$ gates in the complete circuit that computes the nearest ATM location. It shows that in our privacy preserving application of finding the nearest Chase or Wells Fargo ATM in Salt Lake City, the circuit has a total of $2596$ $XOR$ and $854$ $AND$ gates.

\subsection{Server-side and Client-side Cost}

\begin{figure}[t]
\centering
\includegraphics[width=3.25in]{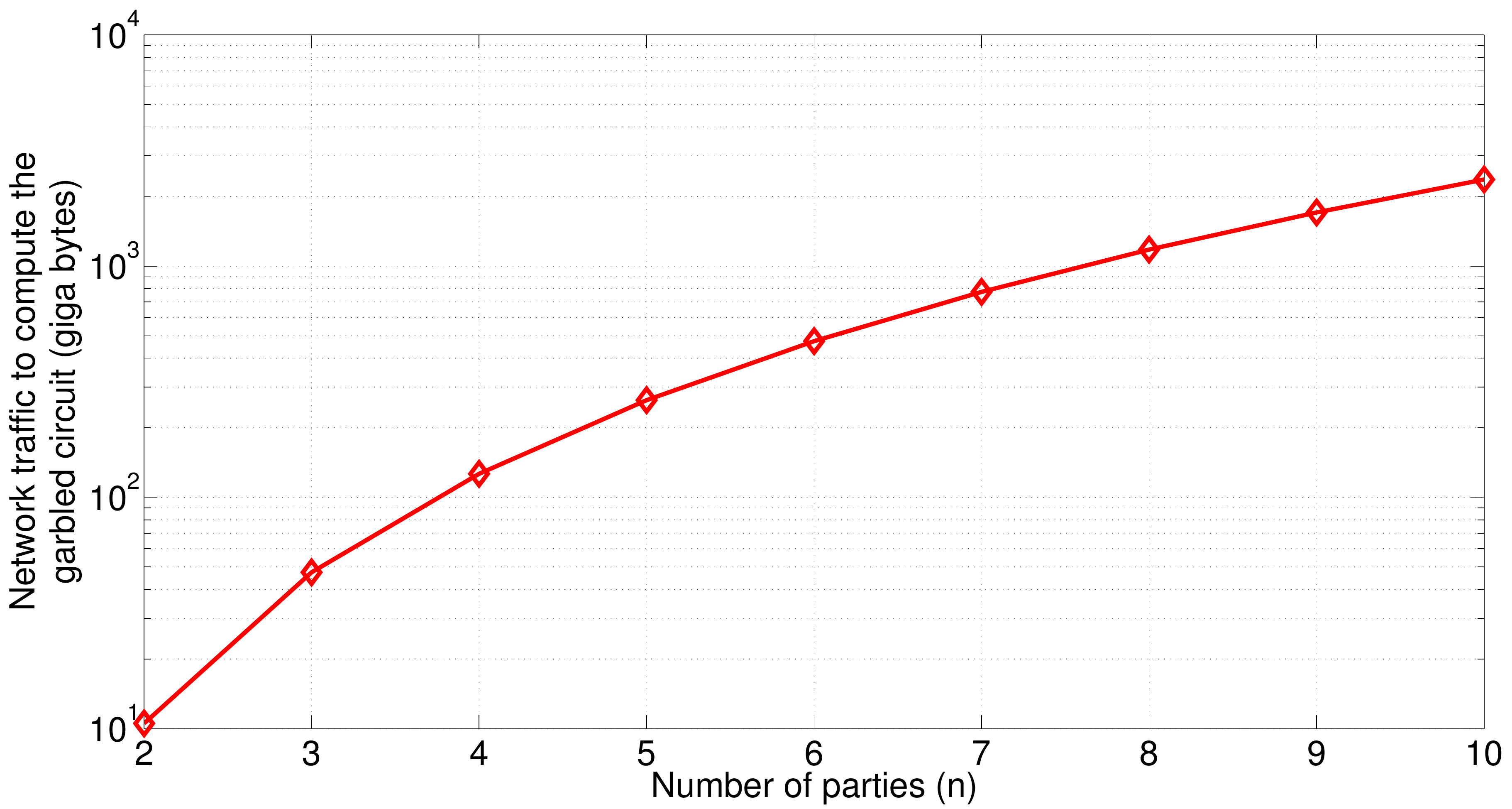}
\caption{Server-side network traffic to construct the garbled circuit for determining the nearest ATM.}
\label{fig_server_side_network_traffic_closest_ATM}
\end{figure}

Figure~\ref{fig_server_side_network_traffic_closest_ATM} shows the network traffic as a function of the number of servers ($n$) involved in the creation of the garbled circuit that can compute the nearest ATM and the corresponding distance in a privacy preserving manner. For example, with $n=4$ servers, the servers exchange a total of $126$ $GB$ of information to create the garbled circuit. This result demonstrates the feasibility of our approach for performing real-world privacy-preserving computations.

\begin{figure}[t]
\centering
\includegraphics[width=3.25in]{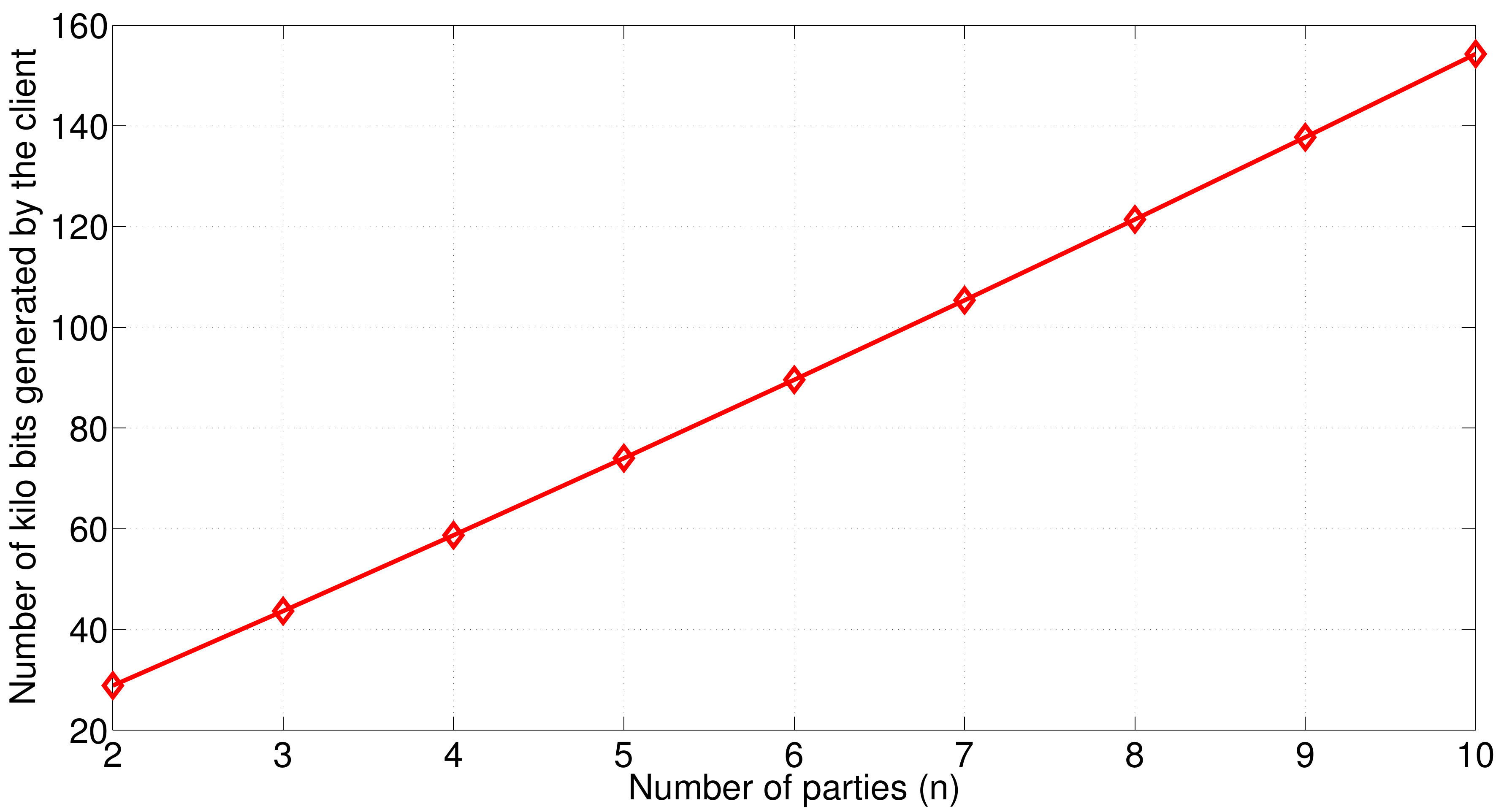}
\caption{Client-side cost to determine the nearest ATM.}
\label{fig_client_side_cost_closest_ATM}
\end{figure}

In order to facilitate the creation of the garbled circuit, and for the evaluation, the mobile client sends -- (i) the seed values to the $n$ servers, and (ii) the garbled values representing the coordinates of the input location to the evaluator. Since the ATM locations are publicly known, they are assumed to be hard-coded in the garbled circuit. i.e., the client is not required to transmit the ATM locations to the servers.

Figure~\ref{fig_client_side_cost_closest_ATM} shows the total number of bits generated by the client to delegate the privacy preserving computation of the nearest ATM. To preserve location privacy, the client exchanges a very small amount of information with the servers -- less than $60$ kilo bits, with $n=4$ servers, for example. Comparing Figure~\ref{fig_server_side_network_traffic_closest_ATM} and Figure~\ref{fig_client_side_cost_closest_ATM}, we note that {\em the client-side cost grows much slowly with the number of servers, in comparison to the server-side cost.}

\section{Related Work}
\label{sec_related_work}

Homomorphic encryption is an approach that enables performing computations directly on the encrypted data, without requiring private decryption keys. For example, in the RSA public key system, the product of two ciphertext messages produces a ciphertext corresponding to the product of the underlying plain text messages \cite{rivest78}. Domingo-Ferrer et al. \cite{domingoFerrer02} present a homomorphic scheme that represents ciphertext as polynomials, allowing both addition and multiplication operations on the underlying plain text; however, in this scheme, multiplication operations drastically increase the size of the cipher text. Recently, fully homomorphic encryption (FHE) schemes (e.g., Gentry et al. \cite{gentry10}) have been proposed, which enable performing any arbitrary computation on encrypted data. However, FHE schemes are currently impractical for cloud computing applications due to extremely large cipher text size (Section~\ref{sec_cost_for_client}).

Yao's garbled circuits have been primarily used in conjunction with oblivious transfer protocols for secure two-party computation \cite{yao82,yao86,lindell09}. Lindell et al. \cite{lindell_survey09} present an excellent survey of secure multiparty computation, along with numerous potential applications, such as privacy preserving data mining, private set intersection, electronic voting and electronic auction. A number of secure two-party and multiparty computation systems have been built over the years (e.g., \cite{henecka10,malkhi04,david08}). Note that in secure multiparty computation systems multiple parties hold private inputs and receive the result of the computation; however, in a secure cloud computing system, such as ours, while multiple parties participate in the creation of garbled circuits, only the client holds private inputs and obtains the result of the computation in garbled form. In our work, we adapt secure multiparty computation protocols \cite{goldreich04,goldreich87,beaver90,rogaway91}, as we have described in Section~\ref{sec_key_changes_to_BMR_Goldreich}, for building a secure and verifiable cloud computing for mobile systems.

Some existing works \cite{burkhart10,damgaard12} use arithmetic circuits for performing homomorphic addition and multiplication operations. However, Boolean circuits are sufficient for our purpose of creating the garbled table entries as it only involves simple binary operations such as XOR and AND (Section~\ref{sec_garbled_circuit_BMR_construction}). Additionally, Boolean circuits are more efficient than arithmetic circuits for homomorphic comparison of numbers, which we have used abundantly in our privacy preserving search application (Section~\ref{sec_manhattan_distance}), since comparison using an arithmetic circuit involves a large number of multiplications and communication rounds among the participants \cite{burkhart10,damgaard12}.


Twin clouds \cite{bugiel11} is a secure cloud computing architecture, in which the client uses a {\em private} cloud for creating garbled circuits and a {\em public} commodity cloud for evaluating them.
Our solution, on the other hand, employs multiple public cloud servers for creating as well as evaluating the garbled circuits. {\em In other words, our solution obviates the requirement of private cloud servers. Furthermore, in twin clouds, the privacy of the client data is lost if the evaluator colludes with the sole/only server that constructs the garbled circuit. In our work, on the other hand, the use of multiple servers for construction of garbled circuits offers greater resistance to collusion (Section 2).}

While FHE schemes currently remain impractical, they, however, offer interesting constructions, such as reusable garbled circuits \cite{goldwasser13} and verifiable computing capabilities that permit a client to verify whether an untrusted server has actually performed the requested computation \cite{gennaro10}. {\em In our proposed system, we enable the client to efficiently verify whether an untrusted server has actually evaluated the garbled circuit, without relying on any FHE scheme.}

Naehrig et al. \cite{naehrig11} present an implementation of a ``somewhat" homomorphic encryption scheme, which is capable of computing simple statistics such as mean, standard deviation and regression, which require only addition operations, and a very few multiplication operations. Another recent work (P4P \cite{duan10}) provides privacy preservation for computations involving summations only. In comparison, our scheme, which is based on Yao's garbled circuits, is not limited to computation of simple statistics or summations only; i.e., our scheme is more generic, and can perform any arbitrary computation.

Carter et al. \cite{carter13} have proposed an atypical secure two party computation system with three participants: Alice, Bob and a Proxy. In their work, Bob is a webserver that creates garbled circuits, and Alice is a mobile device that delegates the task of evaluating the garbled circuits to the Proxy, which is a cloud server. We note that the computation and adversary models in Carter et al.'s work are very different from that of our work. First, in their work, being a secure two party computation system, {\em both Alice and Bob} provide private inputs for the computation that they wish to perform jointly; however, in our secure cloud computing model, {\em only one party, i.e., the mobile client}, provides inputs and obtains result of the computation in garbled form. Second, Cartel et al.'s scheme requires that neither Alice nor Bob can collude with the Proxy; in a sharp contrast, our method preserves the privacy of the client data even if the evaluating server colludes with all but one of the cloud servers that participated in the creation of the garbled circuit. 


\section{Concluding Remarks}
\label{sec_conclusion}

We proposed a novel secure and verifiable cloud computing for mobile system using multiple servers. Our method combines the secure multiparty computation protocol of Goldreich et al. and the garbled circuit design of Beaver et al. with the cryptographically secure pseudorandom number generation method of Blum et al. Our method preserves the privacy of the mobile client's inputs and the results of the computation, even if the evaluator colludes with all but one of the servers that participated in the creation of the garbled circuit. Further, our method can efficiently detect a cheating evaluator that returns arbitrary values as output without performing any computation. We presented an analysis of the server-side and client-side complexity of our system. Using real-world data, we evaluated our system for a privacy preserving search application that locates the nearest bank/ATM from the mobile client. We evaluated the time taken to construct and evaluate a garbled circuit for varying number of servers, and demonstrated the feasibility of our proposed approach.


%



\ifCLASSOPTIONcompsoc
  \section*{Acknowledgments}
\else
  \section*{Acknowledgment}
\fi

Our work is part of the National Science Foundation (NSF) Future Internet Architecture Project, and is supported by NSF under the following grant numbers: CNS-1040689, ECCS-1308208 and CNS-1352880.

\ifCLASSOPTIONcaptionsoff
  \newpage
\fi



%



\bibliographystyle{abbrv}
\bibliography{paper}

\begin{thebibliography}{10}

\bibitem{armbrust10}
M.~Armbrust, A.~Fox, R.~Griffith, A.~D. Joseph, R.~Katz, A.~Konwinski, G.~Lee,
  D.~Patterson, A.~Rabkin, I.~Stoica, and M.~Zaharia.
\newblock A view of cloud computing.
\newblock {\em Commun. ACM}, Apr. 2010.

\bibitem{nist_key_len_recommendations2012}
E.~Barker et~al.
\newblock Recommendation for key management - part 1: General (revision 3).
\newblock {\em NIST Special Publication 800-57}, July 2012.

\bibitem{beaver90}
D.~Beaver, S.~Micali, and P.~Rogaway.
\newblock The round complexity of secure protocols.
\newblock In {\em ACM STOC'90}.

\bibitem{david08}
A.~Ben-David, N.~Nisan, and B.~Pinkas.
\newblock Fairplaymp: A system for secure multi-party computation.
\newblock In {\em ACM CCS}, 2008.

\bibitem{blum86}
L.~Blum, M.~Blum, and M.~Shub.
\newblock A simple unpredictable pseudo random number generator.
\newblock {\em SIAM J. Comput.}, 15(2):364--383, May 1986.

\bibitem{bugiel11}
S.~Bugiel, S.~N\"{u}rnberger, A.-R. Sadeghi, and T.~Schneider.
\newblock Twin clouds: Secure cloud computing with low latency.
\newblock In {\em Proc. CMS}, 2011.

\bibitem{burkhart10}
M.~Burkhart, M.~Strasser, D.~Many, and X.~Dimitropoulos.
\newblock Sepia: Privacy-preserving aggregation of multi-domain network events
  and statistics.
\newblock In {\em Proceedings of the 19th USENIX Conference on Security},
  USENIX Security'10, pages 15--15, Berkeley, CA, USA, 2010. USENIX
  Association.

\bibitem{carter13}
H.~Carter, B.~Mood, P.~Traynor, and K.~Butler.
\newblock Secure outsourced garbled circuit evaluation for mobile devices.
\newblock In {\em USENIX Security}, 2013.

\bibitem{damgaard12}
I.~Damg{\aa}rd, V.~Pastro, N.~Smart, and S.~Zakarias.
\newblock Multiparty computation from somewhat homomorphic encryption.
\newblock In {\em Advances in Cryptology--CRYPTO 2012}, pages 643--662.
  Springer, 2012.

\bibitem{domingoFerrer02}
J.~Domingo-Ferrer.
\newblock A provably secure additive and multiplicative privacy homomorphism.
\newblock In {\em Proc. Int'l Conf. on Information Security}, 2002.

\bibitem{duan10}
Y.~Duan, J.~Canny, and J.~Zhan.
\newblock P4p: Practical large-scale privacy-preserving distributed computation
  robust against malicious users.
\newblock In {\em Proceedings of the 19th USENIX Conference on Security},
  USENIX Security'10, pages 14--14, Berkeley, CA, USA, 2010. USENIX
  Association.

\bibitem{gennaro10}
R.~Gennaro et~al.
\newblock Non interactive verifiable computing: Outsourcing computation to
  untrusted workers.
\newblock In {\em CRYPTO}, 2010.

\bibitem{gentry10}
C.~Gentry.
\newblock Computing arbitrary functions of encrypted data.
\newblock {\em Commun. ACM}, Mar. 2010.

\bibitem{goldreich04}
O.~Goldreich.
\newblock {\em Foundations of Cryptography: Volume 2, Basic Applications}.
\newblock Cambridge University Press, 2004.

\bibitem{goldreich87}
O.~Goldreich, S.~Micali, and A.~Wigderson.
\newblock How to play any mental game.
\newblock In {\em ACM STOC}, 1987.

\bibitem{goldwasser13}
S.~Goldwasser et~al.
\newblock Reusable garbled circuits and succinct functional encryption.
\newblock In {\em ACM STOC'13}.

\bibitem{henecka10}
W.~Henecka et~al.
\newblock Tasty: Tool for automating secure two-party computations.
\newblock In {\em ACM CCS'10}.

\bibitem{huang2012private}
Y.~Huang, D.~Evans, and J.~Katz.
\newblock Private set intersection: Are garbled circuits better than custom
  protocols.
\newblock In {\em Network and Distributed System Security Symposium (NDSS). The
  Internet Society}, 2012.

\bibitem{kolesnikov09}
V.~Kolesnikov, A.-R. Sadeghi, and T.~Schneider.
\newblock Improved garbled circuit building blocks and applications to auctions
  and computing minima.
\newblock In {\em Proceedings of the 8th International Conference on Cryptology
  and Network Security}, CANS '09, pages 1--20, Berlin, Heidelberg, 2009.
  Springer-Verlag.

\bibitem{lindell09}
Y.~Lindell and B.~Pinkas.
\newblock A proof of security of yao's protocol for two-party computation.
\newblock {\em J. Cryptol.}, 22(2):161--188, Apr. 2009.

\bibitem{lindell_survey09}
Y.~Lindell and B.~Pinkas.
\newblock Secure multiparty computation for privacy-preserving data mining.
\newblock {\em Journal of Privacy and Confidentiality}, 1(1), 2009.

\bibitem{malkhi04}
D.~Malkhi, N.~Nisan, B.~Pinkas, and Y.~Sella.
\newblock Fairplay -- a secure two-party computation system.
\newblock In {\em USENIX Security}, 2004.

\bibitem{naehrig11}
M.~Naehrig, K.~Lauter, and V.~Vaikuntanathan.
\newblock Can homomorphic encryption be practical?
\newblock In {\em Proceedings of the 3rd ACM Workshop on Cloud Computing
  Security Workshop}, CCSW '11, pages 113--124, New York, NY, USA, 2011. ACM.

\bibitem{naor01}
M.~Naor and B.~Pinkas.
\newblock Efficient oblivious transfer protocols.
\newblock In {\em ACM SODA}, 2001.

\bibitem{naor05}
M.~Naor and B.~Pinkas.
\newblock Computationally secure oblivious transfer.
\newblock {\em J. Cryptology}, 18(1):1--35, 2005.

\bibitem{rivest78}
R.~L. Rivest, L.~Adleman, and M.~L. Dertouzos.
\newblock On data banks and privacy homomorphisms.
\newblock {\em Foundations of secure computation}, 32(4), 1978.

\bibitem{rogaway91}
P.~Rogaway.
\newblock {\em The round complexity of secure protocols}.
\newblock PhD thesis, MIT, 1991.

\bibitem{schneier95}
B.~Schneier.
\newblock {\em Applied Cryptography (2Nd Ed.): Protocols, Algorithms, and
  Source Code in C}.
\newblock John Wiley \& Sons, Inc., 1995.

\bibitem{boolean_circuits}
S.~Tillich and N.~Smart.
\newblock Circuits of basic functions suitable for {MPC} and {FHE},
  http://www.cs.bris.ac.uk/{R}esearch /{C}ryptography{S}ecurity/{MPC}/.

\bibitem{yao82}
A.~C. Yao.
\newblock Protocols for secure computations.
\newblock In {\em IEEE Computer Society SFCS}, 1982.

\bibitem{yao86}
A.~C.-C. Yao.
\newblock How to generate and exchange secrets.
\newblock In {\em IEEE Computer Society SFCS}, 1986.

\end{thebibliography}

\end{document}